\documentclass[12pt]{article}
\pdfoutput=1
\usepackage{epsf}
\usepackage{amsmath}
\usepackage{graphicx}
\usepackage{color}
\usepackage{psfrag}

\usepackage{epsf}
\usepackage{graphicx,epsfig}
\usepackage{amssymb}
\usepackage{epstopdf}
\usepackage{amsmath}
\usepackage{amsfonts}
\usepackage{latexsym}
\usepackage{color}
\usepackage{xypic}
\usepackage{cancel,slashed}
\usepackage[linktocpage]{hyperref}
\usepackage{cite}

\usepackage{ifpdf}

\def\IZ{\mathbb{Z}}
\def\IR{\mathbb{R}}

\def\s{{\text{sgn}}}

\def\T{{\bf T}}

\def\oh{\frac{1}{2}}
\def\a{{\alpha}}
\def\b{{\beta}}
\def\d{{\delta}}

\def\om{{\omega}}

\def\g{{\gamma}}

\def\p{{\partial}}

\def\CK {{\cal K}}

\def\CF {{\cal F}}
\def\CL {{\cal L}}

\def\re{\mbox{Re }}

\def\be{\begin{equation}}
\def\ee{\end{equation}}
\def\bea{\begin{eqnarray}}
\def\eea{\end{eqnarray}}
\def\bes{\begin{subequations}}
\def\ees{\end{subequations}}
\def\raw{\rightarrow}
\def\Raw{\Rightarrow}

\newcommand{\bmat}{\left(\begin{array}}
\newcommand{\emat}{\end{array}\right)}

\def\pim{{\rm Im\,}}

\def\yzero{\smash{\hbox{$y\kern-4pt\raise1pt\hbox{${}^\circ$}$}}}
\def\p{\partial}
\def\a{\alpha}
\def\b{\beta}
\def\g{\gamma}
\def\d{\delta}
\def\beq{\begin{equation}}
\def\eeq{\end{equation}}
\def\beqa{\begin{eqnarray}}
\def\eeqa{\end{eqnarray}}

\def\om{\omega}

\def\-{\hphantom{-}}

\def\s2{\frac{1}{\sqrt2}}

\def\oh{\frac{1}{2}}

\def\IF{\relax{\rm I\kern-.18em F}}
\def\II{\relax{\rm I\kern-.18em I}}

\def\Dsl{\,\raise.15ex\hbox{/}\mkern-13.5mu D} %this one can be subscripted

\def\IS{{\bf S}}
\def\IR{{\bf R}}
\def\IZ{{\bf Z}}
\def\IX{{\bf X}}
\def\IT{{\bf T}}

\def\re{{\rm Re}\,}

\def\raw{\rightarrow}
\def\Raw{\Rightarrow}

\catcode`\@=11   
\newdimen\@rotdimen
\newbox\@rotbox  

\def\@vspec#1{\special{ps:#1}}%  passes #1 verbatim to the output
\def\@rotstart#1{\@vspec{gsave currentpoint currentpoint translate
   #1 neg exch neg exch translate}}% #1 can be any origin-fixing transformation
\def\@rotfinish{\@vspec{currentpoint grestore moveto}}% gets back in synch 
\def\@rotr#1{\@rotdimen=\ht#1\advance\@rotdimen by\dp#1%
   \hbox to\@rotdimen{\hskip\ht#1\vbox to\wd#1{\@rotstart{90 rotate}%
   \box#1\vss}\hss}\@rotfinish}
\def\@rotl#1{\@rotdimen=\ht#1\advance\@rotdimen by\dp#1%
   \hbox to\@rotdimen{\vbox to\wd#1{\vskip\wd#1\@rotstart{270 rotate}%
   \box#1\vss}\hss}\@rotfinish}%
\def\@rotu#1{\@rotdimen=\ht#1\advance\@rotdimen by\dp#1%
   \hbox to\wd#1{\hskip\wd#1\vbox to\@rotdimen{\vskip\@rotdimen
   \@rotstart{-1 dup scale}\box#1\vss}\hss}\@rotfinish}%
\def\@rotf#1{\hbox to\wd#1{\hskip\wd#1\@rotstart{-1 1 scale}%
   \box#1\hss}\@rotfinish}%
\def\rotate{\@ifnextchar[{\@rotate}{\@rotate[l]}}
\def\@rotate[#1]#2{\setbox\@rotbox=\hbox{#2}\@nameuse{@rot#1}\@rotbox}

\catcode`\@=12
\topmargin
-1.5cm
\textwidth
15.5cm
\textheight
23.5cm
\oddsidemargin
0.7cm
\evensidemargin
1.2cm

\begin{document}

%----------------------------------------------------------------------%
%  numbering equations with section number
%----------------------------------------------------------------------%
\makeatletter
\@addtoreset{equation}{section}
\makeatother
\renewcommand{\theequation}{\thesection.\arabic{equation}}
%----------------------------------------------------------------------%
%  title page
%----------------------------------------------------------------------%
\pagestyle{empty}
%\vspace*{1.0in}
\rightline{ IFT-UAM/CSIC-14-032}
\rightline{MAD-TH-04-01}
\vspace{0.1cm}
\begin{center}
\LARGE{\bf F-term Axion Monodromy Inflation \\[12mm]}
\large{Fernando Marchesano$^1$, Gary Shiu$^{2,3}$, Angel M. Uranga$^1$\\[3mm]}
\footnotesize{$^1$ Instituto de Fisica Te\'orica IFT-UAM/CSIC,\\[-0.3em] 
C/ Nicol\'as Cabrera 13-15, Universidad Aut\'onoma de Madrid, 28049 Madrid, Spain\\[2mm]
$^2$ Department of Physics, University of Wisconsin, Madison, WI 53706, USA \\[2mm]
$^3$ Center for Fundamental Physics and Institute for Advanced Study,\\
Hong Kong University of Science and Technology, Hong Kong} \\[2mm] 

\small{\bf Abstract} \\[5mm]
\end{center}
\begin{center}
\begin{minipage}[h]{16.0cm}

The continuous shift symmetry of axions is at the heart of several realizations of inflationary models. In particular, axion monodromy inflation aims at achieving super-Planckian field ranges for the inflaton in the context of string theory. Despite the elegant underlying principle,  explicit models constructed hitherto are exceedingly complicated.  
We propose a new and better axion monodromy inflationary scenario, where the inflaton potential arises from an F-term. We present several scenarios, where the axion arises from the Kaluza-Klein compactification of higher dimensional gauge fields (or $p$-form potentials) 
in the presence of fluxes and/or torsion homology. The monodromy corresponds to a change in the background fluxes,  and its F-term nature manifests in the existence of domain walls interpolating among flux configurations. Our scenario leads to diverse inflaton potentials, including linear large field behaviour, chaotic inflation, as well as potentials with even higher powers. They provide an elegant set of constructions with properties in the ballpark of the recent BICEP2 observational data on primordial gravitational waves.

\end{minipage}
\end{center}
\newpage
%----------------------------------------------------------------------%
%  Resetting of counters
%----------------------------------------------------------------------%
\setcounter{page}{1}
\pagestyle{plain}
\renewcommand{\thefootnote}{\arabic{footnote}}
\setcounter{footnote}{0}
%----------------------------------------------------------------------%
%  Paper begins
%----------------------------------------------------------------------%

\vspace*{1cm}

\setcounter{tocdepth}{2}

%\vspace*{-1cm}
\tableofcontents

\vspace*{1cm}

\section{Introduction}

The possible first detection of primordial gravitational waves by the BICEP2 collaboration \cite{Ade:2014xna} has opened an exciting  new chapter in observational cosmology. If the observed B-mode
polarization of the cosmic microwave background (CMB)
 is confirmed to be primordial in origin, this discovery
 will be seen as a  watershed for theories of the early universe. 
 Inflation \cite{Guth:1980zm,Starobinsky:1980te,Linde:1981mu} provides a compelling interpretation of this finding, while competing theories which give unobservable tensors (such as \cite{Khoury:2001wf}) are now disfavored. 
 Moreover, 
the large amplitude of gravitational waves as inferred from BICEP2 
(corresponding to a tensor to scalar ratio $r\gg 0.01$),
combined with a simple theoretical argument due to Lyth \cite{Lyth:1996im}, suggests a super-Planckian inflaton field range during inflation. Thus, if the BICEP2 result holds up and is confirmed by other experiments such as Planck, a large class of popular inflationary models -- namely, single, small-field inflation -- are ruled out simply by the observation of a single quantity $r$.\footnote{This is assuming that the observed gravitational waves are produced by vacuum fluctuations. Inflation could also have alternative sources of gravitational waves, e.g., from particle production \cite{Senatore:2011sp, Barnaby:2012xt}. The alternative mechanisms proposed thus far all invoke additional fields. Interestingly, the only known model which has been shown to satisfy the current bounds on non-Gaussianity in the scalar perturbations (Model II in \cite{Barnaby:2012xt}) also involve axions.}

If the observed primordial gravitational waves are indeed generated by vacuum fluctuations during inflation, the implications for quantum gravity are astounding. Not only is the scale of inflation $V^{1/4} = 2.2 \times 10^{16} ~{\rm GeV} \times (r/0.2)^{1/4}$ not far from the Planck scale 
but the  super-Planckian field excursion required for generating the observed large $r$ strongly motivates a UV completion of inflation.
Large-field inflation models
are sensitive to an infinite number of corrections to the inflaton potential which are suppressed by the Planck mass scale. 
Turning this around, understanding how such corrections are controlled in a concrete
framework of quantum gravity, such as string theory, offers a unique opportunity for 
 connecting high scale physics to experiment.

A natural way to suppress the couplings of the inflaton to the heavy degrees of freedom is through
an approximate shift symmetry.
This idea was invoked in the early influential work on natural inflation  \cite{Freese:1990rb}
though to establish that this shift symmetry is respected by Planck scale physics requires embedding it in a quantum theory of gravity.
In string theory, two broad class of ideas to realize large-field inflation with axions have been put forward \cite{Baumann:2014nda}.
 The first class of ideas involves multiple axions in an intricate way \cite{Kim:2004rp,Dimopoulos:2005ac,Grimm:2007hs,Berg:2009tg}, while the second can be implemented with a single axion if there is a non-trivial monodromy in field space \cite{am1,am2}.
In both cases, large field inflation can be achieved without requiring a large axion decay constant $f \gg M_p$ (needed in \cite{Freese:1990rb}) which seems implausible in string theory \cite{Banks:2003sx}.

The axion monodromy idea \cite{am1,am2} is particularly interesting in that the ingredients involved (shift symmetries, branes, and fluxes)
are rather common in string theory. However, it is fair to say that a concrete string theoretical model realizing this idea is far from our sight.
To avoid the supergravity eta-problem, an NS5-brane \footnote{In \cite{Palti:2014kza} a variant of this setup involving (p,q) 7-branes was considered.} (instead of a D5-brane) was introduced in \cite{am2} to break the shift symmetry, while tadpole cancellation requires it to be accompanied by an anti NS5-brane (herefrom $\overline{\rm NS5}$-brane).
To prevent the NS5-$\overline{\rm NS5}$ pair from annihilating perturbatively, they are assumed to be down at different warped throats, wrapping homologous 2-cycles.
It seems exceedingly challenging to construct string compacitifcations (a.k.a. global embeddings) realizing these features.
Furthermore, the backreaction of the NS5-$\overline{\rm NS5}$ pair  seems hard to control \cite{Conlon:2011qp} (see also \cite{Flauger:2009ab}).
In addition to these challenges, it is not clear how inflation ends and why the endpoint is a vacuum with approximately zero vacuum energy.

These models  also present the drawback of containing objects that cannot be analyzed with perturbative techniques, but addressing the eta problem by using the perturbative approximation to the K\"ahler potential, which is certainly questionable. 

This is our main motivation to move on to a new and better class of axion monodromy inflation models, based on F-term potentials (in a clear underlying supersymmetry structure). Our F-term axion monodromy inflation models are perturbative and therefore can be studied with the available effective actions, without the inclusion of  objects that cannot be analyzed with perturbative techniques.  We exploit higher-dimensional gauge fields to generate the shift symmetry of a 4d axion, which subsequently acquires an F-term potential by the introduction of  extra ingredients, in the form of torsion or diverse fluxes, which in fact are already necessarily present in any realistic model addressing moduli stabilization. There is moreover a simple interpretation for the monodromy upon a discrete shift of the axion, required for achieving large field inflation in quantum gravity: it results in an increase of the internal field strength background value for the higher dimensional gauge field (and hence an increase in energy). The framework also admit an elegant explanation for the solution of the eta-problem, in terms of mutual consistency conditions of fluxes and branes.

We describe several classes of models, which lead to a variety of inflaton potentials that can be used as templates for comparison with future data. One class of models describes the axion as a (massive) Wilson line for a higher-dimensional $U(1)$ gauge field. Although most of its ingredients are purely field-theoretical, the need for a UV completion motivates their realization in string theory. Axions from massive Wilson lines for standard gauge fields are present in many string compactifications. Possibly, the most familiar being associated to the discrete Wilson lines present in realistic heterotic string Calabi-Yau compactifications. They are also easily engineered in type II compactifications (and orientifolds thereof), where they arise in the open string D-brane sector, which makes our scenario more amenable to addressing the concerns raised above. 

A further class of models is constructed by going beyond standard gauge fields, and realizing axions as massive `Wilson lines' for generalized $p$-form gauge fields. These are very abundant in type II compactifications with fluxes (either from field strengths, from geometric or non-geometric nature) and/or with torsion homology. We provide several simple realizations, including examples in type IIB orientifolds with NSNS and RR 3-form fluxes; we moreover provide tools to diagnose the presence and properties of such fields, in terms of certain topological couplings in the 4d effective action, or conversely in terms of the properties of domain walls of the 4d theory. Models with massive axions from RR fields provide a simple and very promising setup for inflation, due to their absence from many couplings in the theory, a feature that provides a further rationale for the solution to the eta problem.

This paper is organized as follows. In Section \ref{review} we review the ingredients of axion monodromy (section \ref{generalities}), a sketch of earlier models and our viewpoint on their questionability (section \ref{nogo}).
In Section \ref{open} we introduce the first class of F-term axion monodromy, in terms of massive Wilson line axions: in section \ref{massive} we introduce the basic idea and in section \ref{masstor} we relate their massive nature to torsion homology in the compactification space; in section \ref{hidden-gauge} we find a 4d dual description of the system which displays a 4d gauge symmetry not manifest in the original one, and which connects with the 4d effective theory models in \cite{Kaloper:2008fb,Kaloper:2011jz}, in section \ref{susy-structure} we uncover the F-term nature of the axion potential which we interpret in section \ref{maximum} in terms of 4d domain walls; in section \ref{kinetic} we discuss Wilson lines in the context of string compactifications, describing how enter in the K\"ahler potential and giving explicit examples of massive Wilson lines. In Section \ref{closed-revisited} we construct further classes of F-term axion monodromy models, based on axions from generalized $p$-form fields, both in compactifications with torsion homology (section \ref{torsion-axions}), and in flux compactifications (section \ref{fluxed-axions}). Section \ref{inflation} explores the implications for inflation and its observables, while in Section \ref{conclusions} we present our conclusions. Appendix \ref{more-closed} contains other suggestions for variants of F-term axion monodromy models, illustrating the generality and flexibility of the approach. Appendix \ref{dimred} contains computations regarding the dimensional reduction of massive axions.

\section{Axion monodromy}
\label{review}

\subsection{Generalities}
\label{generalities}

Axions, with their underlying continuous shift symmetry, provide a well-motivated starting point to produce an inflaton potential flat enough throughout a super-Planckian field range. Fields with such shift symmetry are ubiquitous for instance in theories with gauge potentials in extra dimensions,  where the continuous shift symmetry is inherited from the higher dimensional gauge invariance of the theory.  Although the continuous symmetry is in general violated by non-perturbative effects, the surviving discrete shift symmetry keeps the potential under control. The idea is very general, and particular realizations indeed arise  in string theory, where such axions are very generic. For instance they can arise as D-brane positions in toroidal compactifications, or more generically from the KK compactification of higher-dimensional $p$-form fields over $p$-cycles in the internal space. In the latter case the non-perturbative effects arise from euclidean string or brane instantons, and the surviving discrete symmetry is related to charge quantization for these objects.

In order to produce a slow-roll potential over a super-Planckian range, axion monodromy inflation \cite{am1,am2} enriches the scenario by introducing an extra contribution to the potential which increases every time the axion completes a period, hence the term monodromy. Denoting the axion field by $\phi$, the prototypical form of the contribution to the potential in many models is schematically
\be
V\, =\, \sqrt{ L^4+ \langle \phi\rangle^2}
\label{pot}
\ee
This leads to a quadratic potential for small vevs, but behaves linearly for large vevs. In order to ensure that the potential remains flat enough, the model requires an UV completion in which the presence of corrections can be addressed, most notably the infamous eta problem. This motivates its discussion in the framework of string theory. In the following we will discuss why embedding the axion monodromy idea via F-term generated potentials seems the more promising avenue to build successful string theory models of inflation. The reader not interested in this discussion may safely skip to section \ref{open}.

\subsection{F-term axion monodromy}
\label{nogo}

Axion potentials with monodromy arise naturally in string theory models, where extra ingredients like branes and fluxes are present and can lead to additional couplings of the axion. The monodromy often admits an interpretation in terms of the appearance of further induced brane tension, which contribute to the potential energy of the configuration. The structure (\ref{pot}) arises from the (Dirac-Born-Infeld) form of the worldvolume action for the parent branes, on which $\phi$ produces the induced branes. Concrete applications for inflation require the models to ensure that (i) there is no net brane charge induced by the monodromies, as they would violate the tadpole cancellation conditions, (ii) the relevant axion does not mix with geometric moduli, otherwise their appearance in the K\"ahler potential produces an $\eta$-problem, and (iii) the backreaction of the monodromy induced branes should be negligible.

The requirement to address these issues makes present axion monodromy models cumbersome. Naively, one simple setup for axion monodromy is to consider type IIB compactifications with O3/O7-planes, with the axion as the scalar arising from the KK reduction of the NSNS 2-form $B_2$ over a 2-cycle $\Pi_2$ in the compactification space, and introducing a D5-brane wrapped on $\Pi_2$ to generate the monodromy. Specifically, when the axion completes one period, the $B$-field ends up inducing one unit of D3-brane tension. Unfortunately it also induces one unit of D3-brane charge, which would violate the RR tadpole conditions, so this forces to consider D5- anti D5- brane pairs, which for stability must be wrapped on homologous 2-cycles trapped far from each other (e.g. down warped throats). Next, $B$-field axions enter the complex K\"ahler moduli in the following way \cite{gl04}
\be
T_\a \,=\, \frac{3}{4} c_{\a\b\g}v^\b v^\g + \frac{3}{2} i \theta_\a + \frac{3}{8}e^{\phi} c_{\a ab} G^a(G-\bar{G})^b 
\label{Tfield}
\ee
where the 4d scalar fields $(v^\a, \theta_\a, G_a)$ are defined by
\be
J\, =\, \sum_\a v^\a \omega_\a \quad \quad \theta_\a = \int_{\IX^6} C_4 \wedge \omega_\a  \quad \quad C_2 - \tau B_2\, =\, (c^a - \tau b^a) \omega_a \, =\, G^a \omega_a
\ee
with $\omega_\a, \omega_a$ a basis of  harmonic 2-forms such that $[\omega_\a] \in H^2_+(\IX^6, \IZ)$, $[\omega_a] \in H^2_-(\IX^6, \IZ)$, the sign denotes the 2-form parity under the orientifold geometric action and $c_{\a\b\g}$, $c_{\a ab}$ the corresponding triple intersection numbers between these classes. Finally, $\tau = C_0 + i e^{-\phi}$. As a result the combination that enters the K\"ahler potential is
\be
\CK\, =\, - \sum_\a {\rm log\, } \left( {T}_\a + \bar{T}_\a + \frac{3}{2} e^{-\phi} c_{\a ab} b^ab^b \right)
\label{Kb}
\ee
in which $(\theta_\a, c^a)$ are absent. The presence of the $b$-axions in $\CK$ implies that stabilization of the K\"ahler moduli via non-perturbative superpotentials $W\sim e^{-T}$ ultimately leads to an eta problem for the such potential inflation field. 

It has been proposed in \cite{am1,am2}, that this problem can be solved by considering the S-dual systems of NS5-branes (or rather brane-antibrane pairs) coupled to axions from the RR 2-form field. In addition, to prevent strong backreaction of the intermediate flux stretched between the distant NS brane-antibrane pair \cite{Conlon:2011qp}, the whole system has been proposed to be outcast down an overall throat \cite{Flauger:2009ab}. Attempts to improve these combination of axion monodromy ingredients into better motivated string compactifications, for instance in terms of 7-branes, have been proposed in \cite{Palti:2014kza}.

However, a general drawback of all models hitherto is that they use an axion which is known not to appear in the K\"ahler potential at the {\em perturbative} level, but ultimately involve objects (like NS5-branes, or $(p,q)$ 7-branes) not admitting a description at weak coupling;  the untractability of the final system stirs the feeling that the issue of the eta problem in these setups remains to be settled.

In order to be more precise, consider a D7-brane and an anti-D7-brane wrapping a 4-cycle $\Pi_4$ dual to the previous 2-cycle $\Pi_2$. Upon dimensional reduction of their CS actions, the $c$-axion appears in a St\"uckelberg Lagrangian
\be
\CL_{St}^{\rm D7}\, \supset \, (\p c - A_\g + A_\d)^2
\ee
where $A_\g$ is the gauge boson of the D7-brane and $A_\d$ that of the anti-D7-brane. This implies that the combination $U(1)_\g + U(1)_\d$ will remain massless while $U(1)_\g - U(1)_\d$ should become massive. By applying S-duality we obtain that in a system with a (0,1) 7-brane wrapping $\Pi_4$ and its anti-7-brane we have
\be
\CL_{St}^{\rm 7_{(0,1)}}\, \supset \, (\p b - A_\g + A_\d)^2
\label{stuck}
\ee
Again, a St\"uckelberg Lagrangian of this form cannot come from a K\"ahler potential (\ref{Kb}), because by standard field theory arguments $b$ must be protected by a shift symmetry in order to be eaten by a gauge boson, and as a result it cannot appear explicitly in the K\"ahler potential as it happens in (\ref{Kb}). That is, the K\"ahler potential is not invariant under the gauge symmetry of (\ref{stuck})
\beqa
A_\g-A_\d \to A_\g-A_\d+\p\lambda, \quad \quad \quad b\to b+\lambda
\eeqa
and instead, one expects $c$ to appear in $\CK$, in contrast with the naive perturbative expression (\ref{Kb}).

This incompatibility between St\"uckelberg terms and K\"ahler potentials would show up in any model attempting a realization of D-term axion monodromy inflation, as also pointed out in \cite{Palti:2014kza}. This direct difficulty has been avoided in earlier models, for instance in \cite{Palti:2014kza} by removing the problematic $U(1)$ invoking the orientifold projection, and in \cite{am1,am2} by the use of 5-branes  (for which the St\"uckelberg term cancels). However, the earlier analysis illustrates the point that the use of the K\"ahler potential (\ref{Kb}) derived perturbatively together with a brane whose effective action is not well-understood in the same perturbative regime may potentially lead to problems,\footnote{In fact, this is one of the motivations behind deriving $SL(2,Z)$ invariant effective actions in F-theory \cite{GarciaEtxebarria:2012zm}.} and in particular regarding the identification of those axions with genuine shift symmetries. Furthermore the fact that all working models hitherto involve brane-antibrane pairs, and thus hard supersymmetry breaking and possibly difficult-to-control backreactions, makes the very notion of K\"ahler potential unclear in these setups. 

In the present paper we undertake an alternative direction in order to find manifestly consistent (and in fact simpler) realizations of axion monodromy inflation. In particular we would like to consider models of inflation which are compatible with spontaneously (rather than explicit) broken supersymmetry. Then by standard supersymmetry arguments we know that an axionic field cannot enter a D-term potential and enjoy a shift symmetry at the same time.\footnote{Indeed, given a complex scalar $\phi = \phi_1 + i \phi_2$, $\phi_1$ enters a Fayet-Iliopoulous term if and only if $\phi_2$ enters a St\"uckelberg Lagrangian. Then $\phi_2$ must have a shift symmetry such that it cannot enter explicitly in the K\"ahler potential, and so $\phi_1$ must necessarily appear in $\CK$ or otherwise $\phi$ would not have a kinetic term.} We then conclude that the most natural way to realize axion monodromy inflation compatible with a 4d supersymmetric structure is to consider models where axions appear in an F-term potential. As we will discuss in the following, there are plenty of string theory setups where this idea can be realized. Some of the simplest involve axions associated to (massive) Wilson lines, as well as generalization of this idea to higher-degree $p$-form fields.  We gather a further complementary set of ideas along these lines in appendix \ref{more-closed}.

\section{Massive Wilson line axion monodromy}
\label{open}

As already mentioned, axion monodromy models based on closed string axions suffer from several drawbacks.  Since the axion monodromy idea is very general, it is natural to look for alternative realizations in more general setups. In the following, we develop a very general realization, in which the axion is played by a (massive) Wilson line scalar arising from a higher-dimensional gauge field. This is an interesting setup in itself, but also serves as a warmup for Section \ref{closed-revisited}, which discusses the generalization for (massive) axions from higher-dimensional $p$-form fields, which are ubiquitous in type II flux compactifications (and/or in compactifications with torsion homology).

A point of terminology: In the type II language in which we will carry out the discussion, the massive Wilson line axion corresponds to an open string field, so the models could be termed as {\em open string axion monodromy models}. However, the setup is very general and could be applied to heterotic string compactifications, and even beyond string theory, in any model with gauge fields and extra dimensions. From this more general perspective, it is more appropriate to call them {\em massive Wilson line axion monodromy models}. As we will see in section \ref{susy-structure}, the axion potential and monodromy arise from a superpotential, hence they are a particular instance of {\em F-term axion monodromy}, which are explored in further generality in Section \ref{closed-revisited}.

In any gauge theory in extra dimensions there is an obvious set of axionic scalars. These correspond to the Wilson lines of the gauge field along homologically non-trivial 1-cycles of the compactification manifold. The continuous shift symmetry of these scalars arises from the higher-dimensional gauge invariance of the theory, and is therefore extremely robust against corrections; this fact has been extensively exploited in BSM models with extra dimensions, as a protection mechanism for the Higgs mass (building on \cite{Hosotani:1983xw}), and in inflation \cite{Freese:1990rb}. Even though non-perturbative effects violate the continuous shift symmetry, a discrete periodic identification remains, which corresponds to non-trivial identifications from large gauge transformations. 

Wilson lines are also familiar in string theory, and in particular in D-branes. In fact, the first appearance of D-branes in string theory was implicit in the realization that Wilson line scalars (e.g. in toroidal compactifications) can be interpreted as positions in the T-dual circle \cite{Dai:1989ua}. This T-dual picture provides a useful geometrization of the axion properties: its shift symmetry is the translational symmetry along the T-dual $\IS^1$, and its periodicity is the periodicity of the T-dual $\IS^1$.

In order to build axion monodromy inflation models, we need a violation of the shift symmetry (leading to the inflaton potential) and a monodromy which allows the axion to take values in a covering of the basic period (to allow for super-Planckian excursions). We now show that both ingredients are easily reproduced by a simple modification, which we dub `massive Wilson lines'. Moreover, in the D-brane setup, massive Wilson lines admit a simple a very nice interpretation in the T-dual picture.

\subsection{Massive Wilson lines}
\label{massive}

Consider for simplicity a $U(1)$ gauge theory on a compactification space $\Pi$ (note that for D-branes this is in general a subspace in the 6d compactification space), descibed as an $\IS^1$ fibered over some base space $B$. We will consider the regime in which the $\IS^1$ size is much smaller than the length scales in $B$, so that we deal with KK compactification on $\IS^1$ first. If the $\IS^1$ fiber is homologically non-trivial, there is a genuine Wilson line scalar
\beqa
\phi=\int_{\IS^1} A_1
\label{wl}
\eeqa
Equivalently, there is an associated harmonic 1-form $\eta_1$, which locally is of the form $dy$, with $y$ a flat coordinate along $\IS^1$. We may expand the 1-form as $A_1=\phi\,\eta_1$ to obtain a massless scalar $\phi$.
As mentioned earlier, the masslessness of $\phi$ is a consequence of the gauge invariance of the higher-dimensional theory. The dependence of the 4d action on $\phi$ is reduced to derivative terms, or combinations of exponentials $e^{2\pi i \phi}$ violating the continuous shift symmetry, but compatible with the discrete identification. In D-brane models, such terms can arise e.g. from worldsheet instantons, and are in fact present in phenomenologically relevant quantities like Yukawa couplings (see e.g., \cite{Cremades:2003qj}). 

Since the axion decay constants of these fields is sub-Planckian, super-Planckian field ranges require a modification which induces axion monodromies.  In the present setup, we now argue that this modification is fairly simple and amounts to considering the global fibration of $\IS^1$ over $B$ such that the $\IS^1$ is trivial in $H_1(\Pi,\IR)$. Then, even though locally in $B$ there is a Wilson line scalar $\phi$  as in (\ref{wl}), global effects in $B$ remove it from the massless spectrum. Hence we call the 4d field $\phi$ a {\em massive Wilson line}. In the dual language of cohomology, there is still a 1-form $\eta_1$, but it fails to be closed, $d\eta_1\sim\omega_2$, and hence is not harmonic. It is nevertheless still a eigenvector of the Laplacian $\Delta = dd^* + d^*d$ with non-vanishing eigenvalue, and so the expansion $A_1=\phi\,  \eta_1$ yields a 4d field $\phi$ which is no longer massless. It is important to notice that the mass scale for $\phi$ is generated by global effects in $B$, hence it is suppressed by its volume. This singles out this field among the plethora of massive fields in any compactification, in the regime of large volume of $B$.

The trivial nature of $\IS^1$, equivalently the non-closedness of $\eta_1$, implies that changing the 4d vev of $\phi$ corresponds to turning on a non-trivial field strength backgound 
\beqa
F_2=dA_1=\phi \,d\eta_1\sim \phi\, \omega_2
\eeqa
This component of the fields strength should not be regarded as a {\em flux}, since $\omega_2$ is exact, therefore trivial, so $F_2$ integrates to zero over any non-trivial 2-cycle. However, this field strength contributes as a $\phi$-dependent 4d vacuum energy, which will become the inflationary potential. Moreover, its leads to the desired monodromy effect, since circling around a period of $\phi$ implies an overall change in the background $F_2$ (adding one unit of $\omega_2$ to it)\footnote{This mechanism has appeared in the holographic computation of the $\theta$-dependent energy of non-supersymmetric gluodynamics in \cite{Witten:1998uka}, whose multivaluedness was explored in \cite{Dubovsky:2011tu} as a toy model of axion monodromy.}. 

The details of the axion potential depend on the details of the dynamics of the underlying gauge field. A quadratic Yang-Mills kinetic term would lead to a quadratic axion potential, but the former may be just the small field approximation to the complete action. For instance, in the D-brane setup, the DBI action results in a structure (\ref{pot}). We therefore postpone the discussion to the relevant models.

Finally, a last ingredient in the axion monodromy scenario are the non-perturbative effects leading to a superimposed periodic modulation of the axion potential. In the present setup, they arise from instanton effects, which are described by euclidean worldlines of charged particles wrapped on the $\IS^1$, since these are the objects coupling to $\phi$. The periodic nature of this contribution follows from charge quantization of these particles.

\subsection{Massive Wilson lines and torsion homology}
\label{masstor}

There is a slight generalization of the above idea, based on torsion (co)homology. It is possible that the $\IS^1$ is not a boundary, but still remains trivial in $H_1(\Pi,\IR)$. This is the case if it corresponds to a torsion class in $H_1(\Pi,\IZ)$. Namely, the $\IS^1$ is not a boundary of a 2-chain, but some multiple $k$ of its class {\em is} a boundary. Following \cite{Camara:2011jg}, there is a 1-form $\eta_1$ satisfying $d\eta_1=k\,\omega_2$, and the above picture for massive Wilson lines goes through with addition of suitable factors of $k$. 

A simple example is provided by the twisted torus ${\tilde \IT}^3$, described as an $\IS^1$  fibered over $\IT^2$ with first Chern class $k$. This $\tilde \T^3$ does not have three 1-cycles like a $\T^3$ would, but instead two standard 1-cycles and one torsional 1-cycle. This implies that there is a globally well-defined but non-closed 1-form $\eta^1$ such that $d\eta^1 = k\, dx^2 \wedge dx^3$, and so turning a Wilson line along such component implies
\be
A = \phi\, \eta^1 \quad \quad \raw \quad \quad F = dA = \phi\, k \,dx^2 \wedge dx^3
\ee
Upon a monodromy in $\phi$, the system picks up a change in $F_2=k\,dx^2\, dx^3$. Note that the space parametrized by $x^2$, $x^3$, is not a 2-cycle (it is precisely the 2-chain whose boundary is homotopic to $k$ times the fiber $\IS^1$), hence this continuous change of $F_2$ does not violate Dirac quantization. Finally, we note that  the mass of $\phi$ can be shown to scale as $k\, R_1/R_2R_3$ (for small values of $\langle \phi \rangle$, see section \ref{kinetic}), where $R_1$ is the size of the torsion 1-cycle and $R_2$, $R_3$ the sizes of the other two. Hence for $R_1 \ll R_2, R_3$ the massive Wilson line is much lighter than the KK replicas of the true Wilson lines (or other moduli present in D-brane realizations of the model), hence it makes sense to single out this field as the relevant for the dynamics of the system, e.g. during inflation.

\medskip

More in general, we consider a gauge field propagating on 4d spacetime times a $d$-dimensional cycle $\Pi_d$ in the compactification space (for a wrapped D$p$-brane, $d=p-3$). 
The KK reduction in the presence of torsion cycles has been discussed in \cite{Camara:2011jg} (see also \cite{BerasaluceGonzalez:2012vb}), which we adapt to our present purposes.  Recall the isomorphism relation among (co)homology groups \cite{Munkres30,bt24}
\beqa
{\rm Tor} \,H_r(\Pi_d,\IZ) ={\rm Tor}\, H^{r+1}(\Pi_d,\IZ) \; ,\; H_r(\Pi_d,\IZ)=H^{d-r}(\Pi_d,\IZ)
\label{torsion-groups}
\eeqa
In the case of a torsion $\IS^1$ 1-cycle we have the nontrivial torsion classes on $\Pi_d$ include
\beqa
{\rm Tor}\, H_1(\Pi_d,\IZ) ={\rm Tor}\, H^{2}(\Pi_d,\IZ)\, {=}\,  {\rm Tor}\,H_{d-2}(\Pi_d,\IZ)={\rm Tor}\,H^{d-1}(\Pi_d,\IZ)
\label{torsion-mwl}
\eeqa
The first relates the torsion $\IS^1$ to a closed but torsion 2-form $\omega_2$, satisfying  $d\eta_1=k \omega_2$, as already mentioned. The other relations will be useful later on.

\subsubsection{Stringy picture and D-brane T-duals}

The above setup can be considered in gauge field theory, but it is best motivated in string theory, which provides a UV complete picture of the construction. A prototypical setup for massive Wilson lines are the discrete Wilson lines in heterotic compactification on non-simply connected Calabi-Yau threefolds \cite{discWL}. In the following sections, we will rather consider D-brane realizations, which moreover allow for a T-dual geometrization of the monodromy, and the corresponding increase in vacuum energy. 

Consider for example the massive Wilson line on a D7 wrapped on a twisted torus ${\tilde \IT}^3$, and perform a T-duality along the $\IS^1$ fiber. The D7-brane turns into a D6-brane whose position in the T-dual circle is parametrized by $\phi$. The T-dual geometry of the relevant coordinates is actually a trivial $\IT^3$,  and the non-triviality of the twisted torus turns into the presence of $k$  units of NSNS 3-form flux $H_3$ on the T-dual $\IT^3$, $H_3=k\,dx^1dx^2dx^3$. In the absence of the $H_3$ flux, $\phi$ would be (the dual of) a true Wilson line, with a continuous shift symmetry and unit period. The presence of $H_3$ breaks the translational invariance, since some quantities can depend on the NSNS 2-form potential, which we write in a convenient gauge as $B_2=k \,x^1dx^2dx^3$, with $x^1$ a flat coordinate on the $\IS^1$. From the D6-brane perspective, there is a non-trivial pullback $B_2=k \,\phi\, dx^2dx^3$, which leads to a $\phi$-dependent background for ${\cal {F}}_2=F_2+B_2$, which contributes to the vacuum energy (in particular it clearly breaks supersymmetry because the F-term condition for a D6-brane is $\CF=0$). This position-dependent potential in the presence of fluxes was exploited in \cite{osl} to stabilize D7-brane positions, and it is behind the appearance of flux-induced soft terms in the D7-brane effective theory \cite{ciu04,civ}. For a worldsheet description of this and other D-brane monodromies, see \cite{Lawrence:2006ma}.

In these D-brane setups, the instanton effects producing the superimposed periodic modulation of the axion potential are worldsheet instantons wrapped on the 2-chain, and whose boundary wraps ($k$ times) the $\IS^1$. Note that this boundary has the interpretation of an euclidean charged particle worldline, in agreement with our earlier general discussion.

The above T-dual picture can be generalized to fibrations of  $\IS^1$ over a base $B$, and produces a T-dual D7-brane wrapped on $B$ and moving in the T-dual $\IS^1$. This holds for fibrations in which the $\IS^1$ is non-contractible, i.e. does not shrink over any point in $B$. The behaviour for singular fibrations is actually interesting, and can lead to a quadratic axion potential even for large fields, which is easily visualized in a simple T-dual picture. Since the details of this massive Wilson line model differ from those of torsion homology, we offer its discussion in Appendix \ref{taub-nut}.

\subsection{A hidden gauge invariance}
\label{hidden-gauge}

The properties of the massive Wilson line scalar are largely constrained by a hidden gauge invariance of the system. In what follows we describe it for an $\IS^1$ being the generator of $H^1(\Pi,\IZ)=\IZ_k$; the case of a completely trivial $\IS^1$ is recovered by setting $k=1$.  

Recall that we consider a gauge field propagating on 4d spacetime times a $d$-dimensional cycle $\Pi_d$ in the compactification space (for a wrapped D$p$-brane, $d=p-3$), and focus on the properties associated to the massive Wilson line along a torsion 1-cycle. Recall the torsion groups (\ref{torsion-mwl}), which relate the torsion $\IS^1$ 1-cycle to a closed but torsion 2-form $\omega_2$, satisfying  $d\eta_1=k \omega_2$, as already mentioned. We now exploit the second relation in (\ref{torsion-mwl}), which implies that there is a torsion $(d-1)$ form $\sigma_{d-1}$, i.e. such that there is a $(d-2)$ form $\lambda_{d-2}$ satisfying
\beqa
d\lambda_{d-2}=k\,\sigma_{d-1}
\eeqa
We would like to discuss the gauge dynamics, not in terms of the 1-form gauge potential $A_1$, but in terms of the dual $(d+1)$-form gauge potential $A_{d+1}$, and its field strength $F_{d+2}$. Their KK expansion along the above non-harmonic forms reads
\beqa
A_{d+1} = b_2 \wedge \sigma_{d-1} + C_3 \wedge \lambda_{d-2}
\eeqa
At the level of the field strength $F_{d+2}=dA_{d+1}$ we have
\beqa
F_{d+2} = (db_2 -kC_3) \wedge \sigma_{d-1} + dC_3\wedge \lambda_{d-2} 
\eeqa
Plugging it into the dual gauge kinetic term, in the quadratic approximation  $|F_{d+1}|^2$, produces the structure
\beqa
\frac{\mu^2}{{k^2}} \int d^4 x\, |db_2 - kC_3|^2 +\int d^4x |F_4|^2
\label{ks-form}
\eeqa
The notation is $|\alpha_p|^2=\alpha_p \wedge * \alpha_p$, and $F_4=dC_3$, while $\mu^2$ is the eigenvalue of $\lambda$ and $\sigma$ under the Laplacian (see Appendix \ref{dimred} for details). This 4d action has a gauge invariance
\beqa
C_3\to C_3+d\Lambda_2\quad ,\quad b_2\to b_2+k\Lambda_2
\label{gauge-inv}
\eeqa
This gauge invariance is hidden in terms of the standard 1-form gauge potential, but it underlies the remarkable properties of the massive Wilson line. The above action describes a 4d 3-form gauge potential which eats up a 2-form field; both fields acquire a mass $\mu$ in a $p$-form generalization of the Higgs mechanism (see \cite{Quevedo:1996uu}, also \cite{Berasaluce-Gonzalez:2013bba} for a recent general discussion), and its gauge symmetry is very similar to that appearing in the discussion of discrete gauge symmetries \cite{Banks:2010zn}. In fact it underlies the discussion of $\IZ_k$ charged domain walls in \cite{BerasaluceGonzalez:2012zn}. The discussion of axion physics in terms of dual 3-form gauge invariance has been explored, in the context of QCD axions, in \cite{Dvali:2005ws,Dvali:2005an,Dvali:2013cpa}. 
These references also describe the generalization of the above lagrangian for general axion potentials, which in our setup arise when higher order terms in the field strength action are included.

The expression (\ref{ks-form}) for our axion monodromy model fits precisely with the 4d effective theory model in \cite{Kaloper:2011jz,Kaloper:2008fb}, where it was proposed as an inflation model leading to large $r$. The above action was found as dual of an axion with a topological mass term due to coupling to a 4-form field strength, 
\beqa
\int d^4x\, \phi\, F_4
\label{top-mass}
\eeqa
which follows from the mixed term in (\ref{ks-form}), with $d\phi=*_{4d}db_2$.
These references argued that the underlying gauge invariance is very effective in protecting the inflaton potential against UV corrections. We have shown that massive Wilson lines provide a realization of this model, and in the case of string theory provides a UV completion, in fact very necessary to  address the question of sustained slow-roll over a large field range. In particular, our string realization shows that  the higher order corrections in the field-strength present in the DBI action on the D-brane  correct the naive quadratic behaviour of the inflaton potential, and turn it into a linear one. Despite this correction, the potential is compatible with large field inflation, and slow-roll. 

\subsection{Structure in ${\cal{N}}=1$ supersymmetry}
\label{susy-structure}

Even though many of the properties of massive Wilson lines do not rely on supersymmetry, their string embedding motivates their discussion in the context of SUSY. A relevant piece of understanding is the superspace structure of the mass term for the Wilson line axion (or in general, of its potential). We now show that in our setup it always corresponds to an F-term, rather than a D-term. It can be then referred as {\em F-term axion monodromy}, as advanced in the introduction. 

This can be easily shown using the description in the previous section, where the mass term follows from the topological coupling (\ref{top-mass}). The ${\cal{N}}=1$ 3-form supermultiplet was studied in \cite{Groh:2012tf}, with the result that the field strength belongs to a chiral multiplet with the structure
\beqa
S\sim i \theta\theta \, \epsilon^{\mu\nu\rho\sigma} (F_4)_{\mu\nu\rho\sigma}\,+\,\ldots
\eeqa
where dots denote terms of no present interest. Completing the inflaton into a chiral multiplet $\Phi$,  the coupling (\ref{top-mass}) clearly corresponds to the F-term
\beqa
\int d^4x \, d^2\theta\, \Phi\, S
\eeqa
There are several other equivalent ways to support this F-term nature. The increase in energy when the axion circles its underlying basic period is associated to the field strength fluxes this produces; the latter can be equivalently described in terms of the fluxes introduced by domain walls in 4d, coupling to precisely the 3-form $C_3$. As familiar from other setups (e.g. \cite{Gukov:1999ya} for domain walls and superpotentials associated to closed string fluxes) the flux contribution to the energy arises from a superpotential, which in ${\cal {N}}=1$ supersymmetric theories is determined by the BPS domain wall tension.

For example, let us review the case of a D6-brane wrapped on a 3-cycle. In general, a domain wall interpolating between two configurations of a D6-brane, one on a 3-cycle $\Pi_3$ and with worldvolume gauge field $A_1$, and the second on $\Pi_3'$ and gauge field $A_1'$ is constructed as follows. Considering a 4-chain $\Sigma_4$ joining the two 3-cycles (i.e. $\partial{\Sigma_4}=\Pi_3'-\Pi_3$), the domain wall is a D6-brane piece, with worldvolume field strength $F_2$ interpolating between the boundary gauge fields (i.e. $F_2=dA_1'-dA_1$). Using generalized calibrations, the superpotential can be written as \cite{Martucci:2006ij}
\beqa
W\sim \int_{\Sigma_4} (F_2+J_c)^2
\label{supo-d6}
\eeqa
Following \cite{torsion} one can see that this superpotential will be non-trivial in the presence of massive Wilson lines, yielding a quadratic term $\mu^2 \Phi^2$ with $\Phi$ a complexification of the massive Wilson line involving a D6-brane position field. 

The F-term origin of the potential for massive Wilson lines will generalize to other setups in section \ref{closed-revisited}. Its associated gauge symmetry (\ref{gauge-inv}) is ultimately responsible for the decoupling of the axion from the K\"ahler potential. Hence it plays a crucial role in solving the eta problem of inflationary scenarios.

\subsection{Domain wall nucleation and a maximum field range}
\label{maximum}

An important property of F-term axion monodromy is that the flux increase upon shifting the axion is associated to the above described domain walls. In our massive Wilson line models, the 4d 3-form $C_3$ arises from the KK reduction of the higher dimensional dual gauge potential. Hence the objects coupling to $C_3$ are the KK reduction of monopole objects on the $(d+4)$ dimensional gauge theory; since monopoles are real codimension 3, hence span $(d+1)$ dimensions, they can produce 4d domain walls by wrapping the torsion $(d-2)$ cycle $\Pi_{d-2}$ dual to (i.e. with non-zero linking number with) the torsion $\IS^1$. Being sources of $F_2$, they indeed produce the correct jump in the fluxes.

This provides a non-perturbative tunneling mechanism \cite{Kaloper:2011jz}, which migh possibly 
destabilize regions of too large axion values. The flux $F_2$ can disappear by nucleation of the domain wall bubbles, with a given number of flux units less in their interior. This can be relevant to provide a dynamical mechanism cutting off the maximum field range, and thus the number of efolds in inflationary scenarios. Similar domain wall nucleation phenomena and competing unstabilities in the large field regime have been considered in a related (but non-supersymmetric) setup in\cite{Dubovsky:2011tu}.

These domain walls are associated to $\IZ_k$ torsion cycles, so that $k$ minimal domain walls can decay. This may suggest the conclusion that the flux cannot be made to change in $k$ units or more; this is incorrect, as we now explain. In fact, $\IZ_k$ charged domain walls and their decay mechanism were described in \cite{Berasaluce-Gonzalez:2013bba}, in a fairly general class of systems, which includes the present one and those in Section \ref{closed-revisited}. A set of $k$ domain walls, can end on a 4d string given by the same $(d+1)$ dimensional `monopole' objects wrapped on the $(d-1)$-chain $\Sigma_{d-1}$ with $\partial \Sigma_{d-1}=k\Pi_{d-2}$. This means that a set of $k$ infinitely extended  coincident domain walls can nucleate a string loop in their interior, creating a hole which can expand and eat up the domain walls. However the string loop is created with a branch cut filling the hole, across which the axion shifts by $\phi\to \phi+1$, which leads to the same monodromy. Hence a spherical bubble of $k$ domain domain walls, nucleating a configuration with $k$ flux units less, can open up a hole in its surface by nucleating a string, but this does not change the nature of the flux configuration in the interior of the bubble.

Let us illustrate these objects in a concrete example, for instance the D6-brane model in the last section. Just as electric charges under the D6-brane worldvolume correspond to open string endpoints, magnetic sources correspond to boundaries of D4-branes ending on the D6-brane (which are indeed real codimension 3). This in fact agrees with the above picture of the D6-brane domain wall in terms of the induced D4-brane charge it carries. As we said, a domain wall D6-brane wrapped on $\partial{\Sigma_4}=\Pi_3'-\Pi_3$ and with the appropriate gauge field $F_2=dA_1'-dA_1$ will interpolate between two D6-brane configurations $(\Pi_3, dA_1)$ and $(\Pi_3', dA_1')$, with $\Pi_3$ and $\Pi_3'$ homotopic. If $A_1' - A_1 = (n/k) \eta^1$ then the interpolating D6-brane carries the charge of $n$ D4-branes wrapping the torsion 1-cycle $\Pi_1$ of the twisted 3-torus and stretching between $\Pi_3$ and $\Pi_3'$. These D4-branes are then the domain walls that change the worldvolume flux quanta $F_2$. Finally, the 4d string opening up holes in sets of $k$ domain walls is given by a D4-brane wrapped on the 2-chain $\Sigma_2$ associated to $\Pi_1$ and stretching between $\Pi_3$ and $\Pi_3'$.

\subsection{Kinetic terms and the $\eta$ problem}
\label{kinetic}

One important point of the scenario in \cite{am1,am2} is that, to avoid the eta problem, the axion must not appear in the K\"ahler potential. In order to address this question we need to embed massive Wilson lines in a string theory construction. We thus now analyze, in the context of type II models with D-branes, how the open string Wilson lines enter into the K\"ahler potential of a string compactification. We will show that Wilson lines enjoy the same property as many type II closed string axions and do not appear explicitly in the K\"ahler potential, making them viable candidates for inflation fields. 

\subsubsection{Kinetic terms for massless Wilson lines}
\label{kinetic-massless}

The dependence of the K\"ahler potential on open string fields can be extracted by studying the kinetic terms for Wilson line scalars, which arise from the higher-dimensional gauge kinetic term. Ignoring the (possibly $F$-dependent) coefficient corresponding to the brane tension, and expanding to quadratic order, it has the structure
\be
\int_{\IR^{1,3} \times \Pi_p} F\wedge * F
\label{quadF}
\ee
where $\Pi_p$ is the $p$-cycle wrapped by the D-brane. From this,  the coefficient of the 4d kinetic term for the Wilson line scalar is proportional to
\be
\int_{\Pi_p} A \wedge * A 
\label{gen}
\ee
where $A$ is the internal profile of the open string field. If $A$ is a harmonic form we relate it by Poincar\'e duality with a non-trivial one-cycle with minimal length $R_A$, and it is easy to see that the integral (\ref{gen}) roughly scales as ${\rm Vol} (\Pi_p)/R_A^2$. 

\subsubsection{Mixing with closed string moduli}

A closer look at the kinetic term for Wilson line moduli reveals the familiar property that they mix with the closed string moduli. We now study the resulting combinations for the two illustrative cases of D6- and D7-branes. The details in this section are not strictly necessary for the rest of this section, so the impatient reader may wish to jump to section \ref{massive-d6}.

\subsubsection*{D7-branes}

For D7-branes, the wrapped 4-cycle $\Pi_4$ is complex, and so $A = A^{(1,0)} + A^{(0,1)}$ and $* A^{(0,1)} = J \wedge A^{(1,0)}$. Let us expand these forms as
\be
J = \sum_\a v^\a \omega_\a \quad \quad A^{(1,0)}\, =\, w(x) \, \eta 
\ee
where $v^\a$ is the geometrical K\"ahler modulus and $w$ a complex 4d Wilson line scalar. We then have that
\be
\int_{\Pi_4} \eta \wedge *_4 \eta\, =\, \sum_\a v^\a \int_{\Pi_4} \omega_\a \wedge \eta \wedge \bar{\eta} 
\ee
As a result the Wilson line kinetic term in the Einstein frame reads \cite{jl04,jl05}
\be
S_E\, =\, {2\mu_3} \frac{i {\cal C}_\a v^\a}{{\rm Vol}_{\IX_6}}  \int_{\IX^{1,3}} dw \wedge *_4 dw \quad \quad {\rm with} \quad \quad {\cal C}_\a \, =\, \int_{\Pi_4} \omega_\a \wedge \eta \wedge \overline{\eta}
\ee
where $\mu_p = (2\pi)^{-p} \alpha'^{-(p+1)/2}$ and ${\rm Vol}_{\IX_6} = \frac{1}{6}\int_{X_6} J\wedge J \wedge J$. Now, in the presence of such Wilson lines we must redefine our holomorphic K\"ahler variables as \cite{BerasaluceGonzalez:2012vb,Berg:2005ja}
\be
\hat{T}_\a \, =\, T_\a + \frac{1}{2}  {\cal C}_\a w (w-\bar{w}) 
\ee
where $T_\a$ is the holomorphic K\"ahler variable (\ref{Tfield}). Notice that the open string field $w$ and the closed string field $G$ enter this holomorphic variable in a similar way. We finally obtain a K\"ahler potential of the form
\be
\CK\, =\, - \sum_\a {\rm log\, } \left( \hat{T}_\a + \bar{\hat{T}}_\a + \oh {\cal C}_\a  (w - \bar{w})^2 + \dots \right)
\ee
Which generalizes (\ref{Kb}), with the dots stand for extra fields like $G$. Form this expression that the open string axion $\re w$ enjoys a symmetry similar to the RR $c$-field considered in \cite{am1,am2}, and thus in the sense serve equally well for axion monodromy inflation candidates. This is also an explicit manifestation of the conclusions in section \ref{nogo}.

\subsubsection*{D6-branes}

We now consider a D6-brane wrapping a Special Lagrangian 3-cycle $\Pi_3$. Let us describe the Wilson line it as
\be
A\, =\, w (x) \, \eta 
\ee
where $\eta$ is now a real form. We have that the Wilson line kinetic term in the Einstein frame reads \cite{gl11,kw11}
\be
S_E\, =\, \mu_4 \frac{g_s^{-1/4} {\cal H}} {8 {\rm Vol}_{\IX_6}}  \int_{\IX^{1,3}} dw \wedge *_4 d\bar{w} \quad \quad {\rm with} \quad \quad {\cal H} \, =\, \int_{\Pi_3} \eta \wedge *_3 \eta
\label{H}
\ee
In this case we have that any 1-form $\eta$ on $\Pi_3$ can be written as
\be
\eta = [\iota_{X} J]_{\Pi_3}
\ee
for some vector $X$ normal to $\Pi_3$. We also have that
\be
*_{\pi_3} [\iota_{X^I} J]_{\Pi_3} \, =\,  [\iota_{X^I} {\rm Im\, } \Omega]_{\Pi_3}
\ee
We now decompose $\pim \Omega = \sum_\a \re u_\a \beta^\a$, where the integer three-forms $\beta^\a$ are such that 
\be
u_\a\, =\, \int_{\IX_6} ({\rm Re\, } \Omega + i C_3) \wedge \beta^\a
\ee
is either the complexified dilation or a complex structure modulus of the compactification. We can then express
\be
{\cal H} \, =\, \sum_\a {\cal Q}^\a \re u_\a  \quad \quad {\rm with} \quad \quad {\cal Q}^\a \, =\, \int_{\Pi_3} \iota_{X}J \wedge \iota_{X}  \beta^\a
\ee
This time the holomorphic variables read
\be
{N}_\a \, =\, u_\a + \frac{1}{2}  {\cal Q}^\a \xi (\xi-\bar{\xi})
\ee
where
\be
\xi \, =\, w + \iota_{X} J_c
\ee
is the complexification of the Wilson line. The K\"ahler potential takes the form 
\be
\CK\, =\, - \sum_\a {\rm log\, } \left({N}_\a + \bar{{N}}_\a + \oh {\cal Q}^\a (\xi - \bar{\xi})^2 \right)
\ee
so again we have a shift symmetry on the Wilson line. This shift symmetry could have been guessed from the fact that Wilson lines become part of the 3-form potential $A_3$ in the lift to M-theory. 

\subsubsection{Massive Wilson lines, and a D6-brane example}
\label{massive-d6}

For massive Wilson lines the kinetic terms are still given by (\ref{gen}), and so a similar reasoning leads to the absence of these fields in the K\"ahler potential. In order to discuss an explicit D-brane example, let us consider the case of type IIA compactification with geometric fluxes, which are prototypical examples for turning some 1-cycles into torsion classes, and so that mass terms for Wilson line moduli are induced. %The fluxes can also be such that e.g. $dJ= d {\rm Im\, } \Omega = 0$ but $\Omega$ is not harmonic, so some of the moduli $N_\a$ are stabilized by fluxes. 

Consider the following example, given in \cite{torsion} and based on a type IIA toroidal orientifold compactification to 4d. The compactification manifold is given by a six-dimensional twisted torus with metric
\be
ds^2 =  R_1^2(dx^1 - k\, x^6 dx^5)^2 + R_2^2(dx^2 + k\, x^4 dx^6)^2 +  \sum_{i=3}^6 (R_i^2dx^i)^2 
\label{metricIIA}
\ee
as well as O6-planes wrapping $\Pi_3^O = \{x^4=x^5=x^6=0\}$. This metric can be understood in terms of the geometric fluxes (for conventions, see e.g., \cite{Camara:2005dc,Ibanez:2012zz})
\beqa
\om^1_{56} = -k  \quad \quad \om^2_{64}  =k
\label{mfluxes}
\eeqa
which turn the 6-torus into a twisted one, with the torsion cohomology structure structure encoded in
\beqa
d\eta^1= k\, \eta^5\wedge \eta^6\quad ,\quad d\eta^2=k\, \eta^4\wedge \eta^6
\eeqa
where $\eta^i$ denote the 1-forms corresponding to the different directions in the twisted 6-torus. Although we will not need it, the model can become ${\cal {N}}=2$ supersymmetric by the simple addition of a RR 2-form flux
\beqa
F_2 \, = \, k\, (\eta^1 \wedge \eta^4 - \eta^2 \wedge \eta^5)
\label{fluxFIIA}
\eeqa
This is easily checked because the configuration is T-dual (along the directions $1,2,3$) of a type IIB toroidal orientifold compactification (with O3-planes) in a vacuum with ${\cal {N}}=2$ preserving NSNS and RR 3-form fluxes \cite{Kachru:2002he}
\beqa
F_3 & = & k \left(dx^4 \wedge dx^2 \wedge dx^3\,
-\, dx^1 \wedge dx^5 \wedge dx^3\right)
\label{fluxFIIB}\\
H_3 & = & k \left(dx^1 \wedge dx^5 \wedge dx^6\,
-\, dx^4 \wedge dx^2 \wedge dx^6\right)
\label{fluxHIIB}
\eeqa
Consider now a D6-brane on the 3-cycle $\Pi_3 = \{x^2=x^3=x^4=0\}$. Due to the flux $\omega^1_{56}$ in (\ref{mfluxes}), this is a twisted ${\tilde\IT}^3$, with the $\IS^1$ along 1 fibered non-trivially over 5,6. More precisely we have the metric
\be
ds^2 \, = \, R_1^2(dx^1 -k\, x^6 dx^5)^2 + (R_5 dx^5)^2  + (R_6 dx^6)^2
\ee
The Wilson line corresponding to $\eta^1$ belongs to a chiral multiplet $\Phi_1$, which acquires mass due to the superpotential (\ref{supo-d6}), which results in \cite{torsion}
\be
W = \oh m_{\Phi_1}^2 (\Phi_1)^2
\label{wl-mass-d6}
\ee
The value of $m_{\Phi_1}$ can be easily computed in the limit where the vev of $\Phi_1$ is small. In that case the worldvolume flux induced by the massive Wilson line is diluted and we can approximate the flux dependence of the DBI action by the quadratic term (\ref{quadF}). Then by the computations of the appendix \ref{dimtor} we obtain an effective 4d action of the form
\be
\int_{\IX^{1,3}} \left(\p_\mu\phi\, \p^\mu \phi + {\mu^2} \phi^2 \right) d{\rm vol}_{\IX^{1,3}} \quad \quad {\rm with} \quad \quad \mu^2\, =\, \left(\frac{{k} R_1}{R_5R_6}\right)^2
\ee
and so we obtain that for canonically normalised kinetic terms the Wilson line mass is given by
\be
m_{\Phi_1} \, =\,  k\,\frac{R_1}{R_5R_6}
\ee
This result can be easily generalized to any D-brane wrapping a twisted torus, as well as to heterotic compactifications along the lines of \cite{km99}.

For large values of the massive Wilson line vev keeping only the quadratic term (\ref{quadF}) is no longer a good approximation, and one needs to use the full DBI expression to compute the effective 4d action. Following appendix \ref{dimDBI} one obtains
\be
\int_{\IX^{1,3}} \left( \frac{1}{2} Z(\phi) \p_\mu\phi\, \p^\mu \phi +  V(\phi)  \right) d{\rm vol}_{\IX^{1,3}} 
\ee
where 
\beqa
 Z(\phi) &= &\frac{1}{R_1} \sqrt{R_5^2R_6^2 + (2\pi\a')^2 k^2 w^2} \\
 V(\phi) &=  &R_1 \left( \sqrt{R_5^2R_6^2 + (2\pi\a')^2 k^2 w^2}  - R_5 R_6 \right)
 \eeqa
The implications of this kind of potential for inflation will be analyzed in section \ref{inflation}. Postponing the direction of more explicit model building for future work, we turn to exploit the intuitions from massive Wilson lines to build further classes of models, this time based on closed string axions.

\section{F-term axion monodromy and flux compactifications}
\label{closed-revisited}

Let us start with a recap of the key ingredients of the massive Wilson line axion monodromy, in a somewhat more abstract language which will motivate further generalizations. The first is the axion, a scalar with an underlying shift symmetry broken to a discrete periodicity only by non-perturbative effects. The second is an F-term coupling which generates the monodromy and therefore the inflaton potential. The third is the implication that there is a built-in mechanism preventing the appearance of the axion in the K\"ahler potential.

A natural setup to generalize the first ingredient is to consider not standard 1-form gauge fields, but general $p$-form gauge potentials. Namely, we take the 4d axion to be the integral of a $p$-form over a $p$-cycle of the compactification space. These are naturally present in string theory compactifications, and in particular we will focus on 4d compactifications of 10d type II strings, or orientifolds thereof (there are straightforward extensions to compactifications of heterotic strings, M-theory or F-theory). This will result in new {\em closed string} axion monodromy models.
 
The second ingredient is also easy to achieve. Compactifications with background fluxes (either for the field strength of antisymmetric tensor fields, or of geometric or non-geometric nature) lead to superpotentials which can stabilize moduli, in particular the components which correspond to the axions from $p$-forms. As we will show in what follows, the F-term masses for the axions automatically contains the structures discussed in \ref{hidden-gauge}, \ref{susy-structure}. This essentially follows from the fact that the energy increase upon axion monodromy is due to the appearance of extra fluxes, whose contribution to the superpotential can be understood in terms of domain walls. 

Finally, these constructions have a built-in mechanism to prevent the appearance of the axions in the K\"ahler potential, at least in terms of the underlying ${\cal N}=1$ SUSY structure, reproducing the third ingredient. This follows from the considerations of F- vs D-terms in section \ref{nogo}, but also admit a beautiful microscopic interpretation, as follows. Axions can in principle appear in the K\"ahler potential necessarily through the St\"uckelberg mechanism. A typical example in type II models (and orientifolds thereof) is the coupling of closed string moduli as D-terms on the D-brane worldvolume. Now in the presence of fluxes, the axions with F-term masses cannot couple as D-terms to the D-branes because the so-called Freed-Witten consistency conditions forbid the existence of the corresponding wrapped D-branes \cite{Camara:2005dc}. Equivalently, the D-term coupling would render the superpotential not invariant under the D-brane worldvolume gauge transformations. A similar analysis can be carried out when the St\"uckelberg couplings involve vector fields from the closed string sector.

There are several other related ways to make this manifest: The 4d gauge symmetry involved in the St\"uckelberg mechanism is incompatible with the gauge symmetry in the F-term axion mass in section \ref{hidden-gauge}. Equivalently, and translating the couplings into the existence of certain $\IZ_k$-valued soliton configurations, the coexistence of both couplings would lead to the existence of 4d domain walls ending on strings ending on junctions (the former from the F-term couplings and the latter from the D-term couplings), which are simply geometrically not sensible \cite{BerasaluceGonzalez:2012zn}. In the general language of gauged supergravities, the gaugings involved in the D-term and F-term coupling violate the quadratic consistency conditions for the gauging embedding tensor \cite{Aldazabal:2008zza}.

This confirms that F-term axion monodromy, in particular in the context of flux compactifications, is a natural setup to build axion monodromy inflation. In the following we present two basic microscopic mechanisms to implement these ideas in detail. The first, in section \ref{torsion-axions}, is based on the simplest generalization of the massive Wilson line idea to $p$-form fields, and relies on the existence of torsion homology classes in the compactification space. Although these can be present in a standard Calabi-Yau geometry, they can be thought of  as geometric flux compactifications in the spirit of the previous paragraphs. The second, in section \ref{fluxed-axions}, is based on the familiar compactifications with field strength fluxes (although geometric fluxes are occasionally allowed). Concrete model building of examples and details of their inflationary applications (beyond those in section \ref{inflation}) are left for future work.

\subsection{Torsional axion monodromy}
\label{torsion-axions}

Consider a $p$-form gauge potential $C_p$ in a 10d string theory compactified to 4d. The reduction of $C_p$ on the $p$-cycle leads to a 4d scalar $\phi$, which is an axion with a continuous shift symmetry due to the higher-dimensional gauge symmetry of the $p$-form, and which is broken to a discrete periodicity by euclidean wrapped brane instantons. In order to keep $p$ general, we focus on RR fields (in type II models or orientifolds thereof), which moreover have the nice feature of being perturbatively absent from many couplings of the 4d effective action (the analysis is however general and applies to NSNS fields, heterotic models and F- and M-theory compactifications).

Consider the same kind of scalar on a geometry $\IX_6$ with a torsion $p$-cycle. For concreteness, we take
\beqa
{\rm Tor}\,  H_p(\IX_6,\IZ)={\rm Tor}\, H_{5-p}(\IX_6,\IZ)=  {\rm Tor}\, H^{p+1}(\IX_6,\IZ)={\rm Tor}\, H^{6-p}(\IX_6,\IZ)=\IZ_k\nonumber
\eeqa
where we have used the relations (\ref{torsion-groups}). 

The scalar $\phi$ is a (massive) `Wilson line' for the generalized gauge field $C_p$. Let us introduce the generator $(p+1)$-form $\omega_{p+1}$ of $ {\rm Tor} H^{p+1}(\IX_6,\IZ)$, whose $k^{th}$ multiple is exact in terms of some non-closed $p$-form $\eta_p$ as
\beqa
d\eta_p=k\,\omega_{p+1}
\eeqa
we moreover require that $\omega_{p+1}$ is an eigenform of the Laplacian $\Delta = d d^* + d^* d$ with eigenvalue $\mu$, which we assume well below the typical KK scale. Then, the KK reductions of $C_p$ and its field strength $F_{p+1}=dC_p$ along these non-harmonic forms are
\beqa
C_p = \phi\, \eta_p \quad ,\quad F_{p+1} = \phi \,k\,\omega_{p+1}
\eeqa
The fact that the forms are non-harmonic implies that $\phi$ is not a modulus; indeed, changing it changes the background field strength, which contributes to the vacuum energy through its kinetic term. The latter are in general only known at the level of the two-derivative supergravity level, that is quadratic in the field strengths, and results in a quadratic potential for $\phi$.

\medskip

As in the case of massive Wilson lines, the system has a hidden gauge invariance in terms of the dual gauge potential. In order to display it, and recalling (\ref{torsion-groups}) we introduce the dual $(5-p)$- and $(6-p)$-forms $\lambda_{5-p}$, $\sigma_{6-p}$, associated to $H^{6-p}(\IX_6,\IZ)$, satisfying
\beqa
d\lambda_{5-p}=k\,\sigma_{6-p}
\eeqa
We now consider the 10d dual $C_{8-p}$ of $C_p$, and its field strength,  and their expansions along these non-harmonic forms, as in \cite{Camara:2011jg}
\beqa
C_{8-p} & = & b_2\wedge \sigma_{6-p} + C_3\wedge \lambda_{5-p} \nonumber \\
F_{9-p} & = & (db_2 -kC_3) \wedge \sigma_{6-p} + dC_3\wedge \lambda_{5-p} 
\eeqa
The kinetic term, in the quadratic approximation, produces the structure (\ref{ks-form})
\beqa
\frac{\mu^2}{{k^2}} \int d^4 x\, |db_2 + kC_3|^2 +\int d^4x |F_4|^2
\label{ks-form-two}
\eeqa
This enjoys the 4d gauge invariance (\ref{gauge-inv}), which strongly constrains possible corrections to the axion potential. For instance, non-perturbative corrections, violating the continuous shift symmetry of $\phi$, but preserving a discrete periodicity, arise only from $D(p-1)$-brane instantons wrapped on the $p$-cycle, so these small effects are well under control. Furthermore, as argued in the introduction of this section, the D-branes for which $\phi$ couples as a D-term are not consistent in the background, as the spaces they should wrap are no longer cycles, but rather have boundaries due to the torsion.

To display the F-term nature of the above coupling, c.f. section \ref{susy-structure}, we look for the domain walls, given by objects coupling to the 4d 3-form $C_3$. They correspond to a D$(7-p)$-brane wrapped on the $\IZ_k$ torsion $(5-p)$-cycles. As explained in section \ref{maximum}, $k$ domain walls can end on a 4d string, given by the D$(7-p)$-brane on the $(6-p)$-chain whose boundary is  $k$ times the torsion $(5-p)$-cycle.

\subsection{Fluxed axion monodromy}
\label{fluxed-axions}

As announced in the introduction of this section, models of axion monodromy are realized in standard compactifications with field strength fluxes. In what follows we present a few illustrative classes of such realizations.

Before getting started, we make a subtle but interesting point. It is worthwhile to mention that these models of axion monodromy are also based on a notion of $\IZ_k$ torsion, but realized not in terms of (co)homology, but rather of K-theory, which is the appropriate tool to characterize D-branes charges and RR fluxes \cite{Witten:1998cd,Moore:1999gb} (or some suitable generalization of K-theory). In particular, in the presence of fluxes certain brane wrappings, allowed by integer-valued homology, are actually not possible; conversely, certain branes wrapped on homologically non-trivial cycles actually carry no conserved charge and can decay through suitable flux-induced processes. The physical picture for these phenomena, based on the so-called Freed-Witten condition,\footnote{Actually \cite{Freed:1999vc} considered the case of torsion $H_3$, and the physical picture for general $H_3$ appeared in \cite{Maldacena:2001xj}.} was proposed in \cite{Maldacena:2001xj} (see also \cite{Evslin:2001cj,Evslin:2002sa,Evslin:2003hd,Evslin:2004vs,Collinucci:2006ug,Evslin:2007ti,Evslin:2007au}). These effects were dubbed {\em flux catalysis} in \cite{BerasaluceGonzalez:2012zn}, where they were associated to certain topological couplings in the 4d action. As shown below, for domain walls they have precisely the structure (\ref{top-mass}), and are part of the F-term flux superpotential. As already mentioned, the disappearance of certain brane wrappings underlies the absence of the axion from D-terms in the effective action.

\subsubsection{Type IIB orientifolds with NSNS and RR 3-form fluxes}

We start with the prototypical class of type IIB Calabi-Yau orientifold compactifications, with O3-planes, and with NSNS and RR 3-form fluxes \cite{Dasgupta:1999ss,Giddings:2001yu} (F-theory generalizations can be discussed similarly).  We introduce a symplectic basis of 3-cycles $\{A_i,B^i\}$, with $A_i\cdot B^j=\delta_i^j$ and introduce the flux quanta
\beqa
\int_{A_i} F_3=n_i \quad,\quad  \int_{B_i} F_3= n_i' \quad ,\quad \int_{A_i} H_3=m_i \quad,\quad  \int_{B_i} H_3= m_i' 
\eeqa
The 4d flux superpotential is \cite{Gukov:1999ya}
\beqa
W=\int_{\IX_6} (F_3-\tau H_3)\wedge \Omega = \int_{\IX_6} \big[\, (F_3-\frac i{g_s} H_3) \, -C_0\, H_3\big]\wedge \Omega
\eeqa
where $\tau=C_0+i/g_s$, with $C_0$ the 10d IIB axion and $g_s$ the string coupling.

We focus on the axion $\phi=C_0$, which has an underlying shift symmetry (reduced to a discrete periodicity by D$(-1)$-brane instantons). Due to the superpotential, there is a non-trivial monodromy $C_0\to C_0+1$ on the fluxes as
\beqa
F_3\to F_3+H_3 \; , \; {\rm i.e.}\;\; (n_i,n_i';m_i,m_i')\to (n_i-m_i,n_i'-m_i';m_i,m_i')
\eeqa
The increase in the tension (or jump in the fluxes) is associated to a domain wall given by a D5-brane wrapped on the 3-cycle Poincare dual to $[H_3]$, namely
\beqa
\Pi_{\rm d.w.}= \frac{m_i'}m A_i\, -\, \frac{m_i}m B_i
\eeqa
with $m={\rm g.c.d}(m_i,m_i')$.
Note that the integral of $H_3$ over this D5-vanishes, so it satisfies the Freed-Witten consistency conditions. These domain walls couple to the 4d 3-form 
relevant in the alternative description (\ref{top-mass}) of the F-term stabilization of the axion. Indeed, the topological coupling to the 4d 4-form field strength arises from the KK reduction of the 10d Chern-Simons coupling
\beqa
\int_{10d} C_0 H_3\wedge F_7 \quad 
\eeqa
where $F_7$ is the 10d dual of the RR 3-form field strength.  Defining
\beqa
F_4=\int_{\Pi_{\rm d.w.}} F_7
\eeqa
we have a 4d coupling\footnote{For a similar computation in the context of M-theory in $G_2$ manifolds see \cite{Beasley:2002db}.}
\beqa
\int_{4d} C_0 \int_{\IX_6 }H_3\wedge F_7 =\int_{4d} C_0 \int_{\Pi_{\rm d.w.}} F_7=\int_{4d} C_0 F_4
\eeqa
The domain walls describe above are $\IZ_k$ valued, as mentioned above. Specifically, in this case we have $k=\sum_{i}[(m_i)^2+(m_i')^2]/m$ with $m={\rm g.c.d}(m_i,m_i')$ \cite{BerasaluceGonzalez:2012zn}. A 4d string formed by a D7-brane fully wrapped on $\IX_6$ must be bounded by $k$ D5-branes on $\Pi_{\rm d.w.}$, due to the non-trivial $H_3$ flux on its worldvolume.
 
\medskip

In the setup of type IIB with NSNS and RR 3-form fluxes, the only other axions preserved by the O3-projection correspond to the integrals of the RR 4-form along orientifold even 4-cycles. These belong to K\"ahler moduli, which are not stabilized by fluxes. Although non-perturbative effects (euclidean D3-brane instantons or non-perturbative gauge dynamics on D7-branes) do produce superpotentials for these moduli (as exploited for their stabilization \cite{Kachru:2003aw}), they do not lead to monodromy and do not allow for super-Planckian field ranges. Flux superpotentials for these moduli can be generated with the introduction of non-geometric fluxes, but these are poorly understood in non-toroidal geometries. We therefore prefer to turn to type IIA constructions.
 
\subsubsection{Type IIA orientifolds with NSNS and RR and geometric fluxes}

It is easy to construct other examples of axion monodromy models based on flux compactifications in type IIA. For compactifications with NSNS and RR field strength fluxes, and geometric fluxes (characterized in terms of a non-closed forms $J_c$ and $\Omega$), the relevant superpotential terms are \cite{Gukov99,tr99,gl04a}
\beqa
\int_{\IX_6} e^{J_c}\wedge (F_0+F_2+F_4)\quad , \quad \int_{\IX_6} ({\rm Re}\, \Omega+iC_3)  \wedge dJ_c
\label{supo-iia}
\eeqa
where $J_c=J+iB$, and so $dJ_c=dJ+iH_3$.

The axions surviving the orientifold projection by O6-planes are: NSNS axions from $B_2$ on orientifold odd 2-cycles, i.e. in $H_{1,1}^-(\IX_6)$, and RR axions from $C_3$ on orientifold even 2-cycles, i.e. $H_{2,1}^+(\IX_6)$. This opens up a whole model building industry, which we leave for future work. We restrict ourselves to presenting an illustrative example of how the 4d flux stabilization technology can be exploited to generate more general  axion monodromy potentials.

For illustration, we consider a simple example, based on an NSNS axion. Consider for simplicity a Calabi-Yau threefold with $h_{1,1}=1$, so there is one K\"ahler modulus $T=\int_{\Sigma_2} (J+iB_2)$ and we focus on the axion $\phi={\rm Im}\, T$. This field can acquire an F-term superpotential from the different RR field strength fluxes $F_0$, $F_2$ and $F_4$ (while other geometric or NSNS fluxes may be present e.g. to achieve a Minkowski and/or supersymmetric vacua, see \cite{Camara:2005dc,Ibanez:2012zz} for examples). The first superpotential in (\ref{supo-iia}) takes the form
\beqa
W=\, e\, T\, -\,q\, T^2 \, +\, m\, T^3
\eeqa
with 
\beqa
m\sim F_0 \; ,\; q\sim \int_{\Sigma_2} F_2\; ,\; e\sim \int_{\Sigma_4} F_4
\eeqa
where $\Sigma_2$, $\Sigma_4$ are the 2-cycle and its dual 4-cycle.

An interesting novelty of this model is that it shows that the axion can acquire potentials which are other than linear or quadratic. In the microscopic 10d description, this follows because the introduction of the Romans mass parameter $F_0$ in the IIA theory makes the field strengths pick up terms depending on powers of $B_2$.

The structure of domain walls, their $\IZ_k$ nature, and their relation to 10d Chern-Simons couplings has been analyzed in section 6.4. of \cite{BerasaluceGonzalez:2012zn}), so we refrain from repeating the general case here. For illustration, we consider the simple case of just turning on $F_0=k$.
There is a 10d Chern-Simons coupling
\beqa
\int_{10d} F_0\, B_2 \wedge F_8
\eeqa
where $B_2$ is the NSNS 2-form and $F_8$ is the field strength dual to that of the RR 1-form $C_1$. Upon KK reduction, we have a 4d coupling
\beqa
k \int_{4d} \phi F_4\quad  \;{\rm with} \; \quad \phi=\int_{\Sigma_2} B_2\quad ,\quad F_4=\int_{\Sigma_4} F_8
\eeqa
reproducing the coupling (\ref{top-mass}). The objects that couple to the 3-form are domain walls charged under the 4d 3-form $\int_{\Sigma_4} C_7$, hence given by D6-branes on $\Sigma_4$. They can annihilate in sets of $k$ by ending on a 4d string, which corresponds to an NS5-brane on $\Sigma_4$, which has a Freed-Witten-like anomaly due to $F_0$.

\section{Applications to inflation}
\label{inflation}

As we have argued in Section \ref{nogo}, earlier works on monodromy inflation 
may secretly suffer from the infamous eta problem, even though this problem may not be apparent from the perturbative effective action of string theory.
In this work, we proposed a new F-term axion monodromy inflationary scenario which evades this problem, and  presented several realizations (via, e.g., massive Wilson lines, torsion homology, flux compactifications, etc) of our ideas.
The forms of the inflaton Lagrangian given in the previous sections are by no means exhaustive.
 Here we briefly discuss the phenomenological features of these models, leaving a more detailed study and further model building to future work.

To deduce the inflationary dynamics and cosmological signatures of our models, we need to know not only the potential (discussed in previous sections) but also the kinetic term for the inflaton. 
We refer the readers to Appendix \ref{dimred} for detailed formulae. Already from the form of the effective actions, we can make some preliminary statements about the implications of our results to large field inflation.

Let us start with our warmup scenario which involves massive Wilson lines. For the twisted tori example in Section 3,
 the action up to two derivatives order takes this form:
 \begin{equation}
 S = - \mu_6 g_s^{-1} \int dx^4 \sqrt{-g_4} \left( \frac{1}{2} Z(w) (\partial w)^2 + V(w) \right)
 \end{equation}
 where
 \begin{equation}
 Z(w) = \frac{1}{R_1} \sqrt{R_5^2R_6^2 + (2\pi\a')^2 k^2 w^2} 
 \end{equation}
and
 \begin{equation}
 V(w) =  R_1 \sqrt{R_5^2R_6^2 + (2\pi\a')^2 k^2 w^2}  - R_1 R_5 R_6
 \end{equation}
Note the similarity of the above action with that of the model in \cite{am1}, apart from the $-R_1R_5R_6$ term in the potential $V(w)$ that arises from taking into account the negative tension of the orientifold planes. This difference is indicative that in contrast to previous works, we have a well-defined endpoint of inflation (when $w=0$). Since this difference lies in the potential and not the kinetic term, we can use the same field redefinition as in \cite{am1} to bring the action to the canonical form:
\begin{equation}
S = \int d^4 x \sqrt{-g_4} \left( - \frac{1}{2} (\partial \phi)^2 - V(\phi) \right)
\end{equation}
with
\begin{equation}
V(\phi) = - \mu_6 g_s ^{-1} \left( R_1 \sqrt{R_5^2 R_6^2 + (2 \pi \alpha')^2 k^2 w^2 (\phi)} - R_1 R_5 R_6 \right)
\end{equation}
where $w(\phi)$ is to be read as the inverse function of the following field redefinition:
\begin{equation}
\phi \sim w \left[ F^2_{1,\frac{1}{2},\frac{3}{4},\frac{3}{2} } \left(-\frac{(2 \pi \alpha')^2 k^2}{R_5^2R_6^2} w^2 \right) + 2 \left( 1+ \frac{(2 \pi \alpha')^2 k^2}{R_5^2R_6^2} w^2 \right)^{1/4} \right]
\end{equation}
Here $F^2_{p,q,r,s} (x)$ is a hypergeometric function.
 In general, the inversion can only be done numerically, though analytic expressions for the inflaton action can be obtained in the limiting regimes, $w << w_{\rm crit}$ or $w>>w_{\rm crit}$, where 
\begin{equation}
w_{\rm crit} = \frac{R_5 R_6}{2 \pi \alpha' k}
\end{equation}
The potential $V(\phi)$ interpolates between a $\phi^2$ potential (small $w$ limit) and a $\phi^{2/3}$ potential (large $w$ limit)  
which gives $r \sim 0.14$ and $r \sim 0.04$ respectively (at 60 e-folds before the end of inflation).
In \cite{am1}, the $\phi^2$ potential was ``ruled out" theoretically based on moduli stabilization consideration (having the simple mechanism \cite{Silverstein:2007ac} and its further simplifications \cite{Haque:2008jz, Danielsson:2009ff, Danielsson:2011au} in mind, see \cite{Gur-Ari:2013sba} for a recent discussion). In the massive Wilson lines  inflationary scenario where moduli stabilization and the inflaton potential can be generated through different means, the requirement on the inflaton potential to not destabilize the background geometry is expected to work differently. Furthermore, one can potentially construct a variety of potentials (and hence a potentially wider range of theoretically allowed values of $r$) by considering other geometries (other than the above twisted tori) with non-closed 1-forms.
Because the idea of massive Wilson lines is rather general, one can examine such model building possibilities not only in type II string theory, but also in heterotic string, M or F-theory, or even simply gauge theories in higher dimensions. 
We defer a detailed study of these interesting new possibilities to future work.

Let us now turn to other realizations of F-term axion monodromy inflation. 
Torsional axion monodromy and fluxed axion monodromy are even more interesting not only because of their wide applicability in different formulations of string theory,
but also the flexibilities and scopes they offer in model building. These scenarios
share some similarities so we describe them together here. To lowest order in the canonically normalized inflaton field $\phi$, the action in either case takes this simple chaotic inflation form (see Appendix \ref{dimred}):
\begin{equation}
S = \int d^4 x \sqrt{-g_4} \left( -\frac{1}{2} (\partial \phi)^2 - \frac{1}{2} \rho^2 M_P^2 \phi^2  + \dots \right)
\end{equation}
where detailed forms of $\rho$ for torsional axion monodromy and fluxed axion mondoromy respectively are given in the Appendix. Thus, these models provide a UV completion of chaotic inflation in the context of string theory. Moreover, one expects terms higher order  in $\phi$ to be generated. For example, the superpotential for fluxed axion monodromy can take the form $W =e T = q T^2 + m T^3$ (where the axion $\phi={\rm Im} T$). In principle this leads to a polynomial inflaton potential of the form $V(\phi) = \sum_{n} c_n \phi^{n}$, with leading power $n\geq2$. Moreover, $\alpha'$ corrections to the effective action which are generically present, such as the $F^4$ term in the effective action of type II string theory recently computed in \cite{Liu:2013dna} can lead to higher order terms in $\phi$. Such higher order terms can help lift the value of $r$ above the predicted value for chaotic inflation $r \sim 0.14$ (at $60$ e-folds before inflation ends), a welcome feature if the high central value of $r$ measured by BICEP2 continues to hold up. Again, we leave detailed model building and thorough analysis of the signatures of these interesting new possibilities for future work.

\section{Conclusions}
\label{conclusions}

Motivated by the recent observational data of the BICEP2 collaboration, we have proposed a new class of string theory models realizing the idea of axion monodromy inflation\footnote{Monodromies are ubiquitous in string theory. In addition to our F-term axion monodromy inflationary scenario, they have been employed differently to realize chain inflation \cite{Chialva:2008xh} and a variant of D-brane inflation \cite{Shlaer:2012by} in string theory.} . The main new ingredient of our constructions is that the axion potential is generated by an F-term perturbative coupling, as opposed to D-term-like potentials, or hard supersymmetry breaking of previous proposals. As we have discussed, 
axion monodromy models based on perturbative F-term potentials naturally avoid such $\eta$ problem by involving axions that do not appear in the K\"ahler potential due to an underlying gauge invariance. This framework moreover allows to construct elegant and simple inflationary models which are compatible with (spontaneously broken) supersymmetry, with a well-defined endpoint of inflation. In fact, to our knowledge, the axion monodromy models presented here are the first ones compatible with a low energy supergravity limit of string theory.

We have illustrated the above features with a series of explicit examples, the simplest being those based on massive Wilson lines.\footnote{For models of D-brane inflation based on standard Wilson lines see e.g. \cite{Avgoustidis:2006zp}.} While our explicit examples of massive Wilson lines are based on D6-brane models, one can implement this construction for any kind of type II or heterotic compactification, the main requirement being that the manifold where gauge degrees of freedom live contains torsional one-cycles. In general, manifolds with torsional homology and/or the presence of background fluxes generate F-term potentials for axionic fields, providing further realizations of the F-term axion monodromy scenario. While one may obtain several kinds of different models they all share certain common features, like being related to the effective theory models of \cite{Kaloper:2008fb,Kaloper:2011jz,Kaloper:2014zba} via an underlying gauge invariance not manifest at first sight. This allows to address several of the points considered in \cite{Kaloper:2008fb,Kaloper:2011jz} from the vantage point of explicit string theory constructions (which is necessary for inflation models with super-Plankian field range). In particular, we have discussed the non-perturbative tunnelling mechanism of \cite{Kaloper:2011jz}, relevant for putting an upper bound on the inflation range, in terms of 4d domain wall bubble nucleation.

While we have not attempted to perform an exhaustive search of F-term axion monodromy models, we have considered a few simple examples that provide interesting inflaton potentials, already showing the possibilities of this new class of models. We have in particular reproduced the models of chaotic inflation and the DBI-like potentials of \cite{am1}, as well as discussed simple possibilities to obtain potentials of the form $V(\phi) = \sum_{n} c_n \phi^{n}$, with $n\geq2$. We expect many more interesting models of inflation to arise from this framework, partly due to the knowledge on perturbative superpotentials that has been developed for the moduli stabilization program carried out during the past decade. In this respect, it is important to note that most of the mechanisms for moduli stabilisation developed so far are based on F-term potentials generated by the presence of background fluxes. This may pose an interesting challenge in the construction of viable string models of inflation, in the sense that all the other compactifications moduli must be stabilized at scale higher than the inflaton mass $10^{13}$ GeV (see e.g. \cite{bww14} for a recent estimate in a specific scenario).  Absent other new ideas of moduli stabilization,
one may have to arrange the F-term potential affecting the inflaton to be hierarchically suppressed compared to the potential affecting other moduli, via considering different flux densities, and/or large warping factors, etc.
 Turning this around, observational data has given us indirect hints on the nature of moduli stabilization which lies at the very heart of string theory, or at the very least motivates us to consider new scenarios.
  In any event, we foresee a very exciting era for string cosmology, where deep theoretical ideas in fundamental physics meet experimental data.

\newpage

\section*{Acknowledgments}

We thank Pablo G. C\'amara, Luis Ib\'a\~nez, Albion Lawrence, Eran Palti, Sam Wong for discussions, and David Andriot, Mikel Berasaluce-Gonz\'alez, Daniel Junghans, Guillermo Ramirez, Pablo Soler, Wieland Staessens, Fang Ye for collaborations in related topics. 
FM and AU are partially supported by the grants  FPA2012-32828 from the MINECO, the ERC Advanced Grant SPLE under contract ERC-2012-ADG-20120216-320421 and the grant SEV-2012-0249 of the ``Centro de Excelencia Severo Ochoa" Programme. F.M. is supported by the Ram\'on y Cajal programme through the grant RYC-2009-05096 and by the REA grant agreement PCIG10-GA-2011-304023 from the People Programme of FP7 (Marie Curie Action). GS is supported in part by the DOE grant DE-FG-02-95ER40896, and would like to thank IFT-Madrid for their hospitality during his visit in January 2014, when this project was initiated.

\newpage

\appendix

\section{Some variants of F-term axion monodromy}
\label{more-closed}

\subsection{Massive Wilson lines in Taub-NUT}
\label{taub-nut}

In this section we develop a simple non-compact picture of an open string axion monodromy model. It admits a T-dual picture in terms of brane motion with a simple geometric picture of the monodromy and energy increase in terms or stretched branes, but which differs significantly from the image in Figure 1 in \cite{am2}.

Consider the case that the massive Wilson line is associated to an $\IS^1$ which is trivial in homology because it is contractible, namely there is a locus in $B$ at which the $\IS^1$ shrinks to zero size. This situation is familiar in elliptic fibrations, where 1-cycles of the elliptic fiber degenerate at special loci. But possibly the simplest model of an $\IS^1$ degeneration is provided by the Taub-NUT geometry, with describes an $\IS^1$ non-trivially fibered over $\IR^3$, asymptotically of the form $\IR^3\times \IS^1$, and with the fiber degenerating at the origin in $\IR^3$. In this geometry, there is a 1-form $\eta_1$ asymptotically of the form $\eta_1=dy$, with $y$ a flat coordinate along $\IS^1$, but which globally fails to be closed. In other words, there is a 2-form $\omega_2$ which asymptotically is of the form $\omega_2=d\eta_1$. The 2-form $\omega_2$ is localized near the $\IS^1$ degeneration at the Taub-NUT center, therefore the monodromy as $\phi$ circles its basic period is an increase in the field strength $F_2$ near the Taub-NUT core.

This model differs from those in section \ref{open} in that the holonomy of the 2-form $\omega_2$ supported on the Taub-NUT center is actually not  $dy$, but tends to it asymptotically far away from the center. A related difference is that the monodromy induced flux $F_2=\phi \omega_2$ leads to the net appearance of induced D3-brane charge, whose tension grows quadratically with $\phi$, even for large fields. This last feature receives an interesting interpretation in the T-dual picture.

\begin{figure}[!ht]
\begin{center}
\includegraphics[scale=.30]{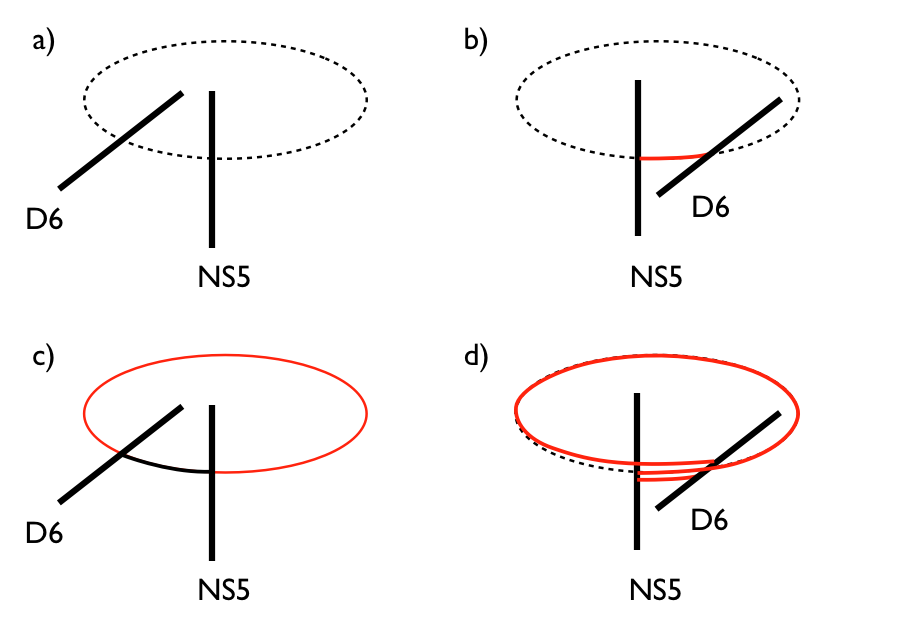}
\caption{\small a) The T-dual of a D7-brane on a Taub-NUT is a D6- and an NS5-branes at points in the T-dual circle. b) As the D6-brane crosses the NS5-brane, an stretched D4-brane is created. c) Moving to larger values of $\phi$ the D6-brane returns close to the NS5-brane. d) Creation of additional D4-branes in each crossing implies the monodromy in $\phi$. The additional brane creation in each crossing makes this construction differ  from Figure 1 in \cite{am2}.
\label{fig:hw}}
\end{center}
\end{figure}

Consider the massive Wilson line on a D7-brane wrapped on a Taub-NUT geometry, and perform a T-duality along the $\IS^1$ fiber. The D7-brane turns into a D6-brane whose position in the T-dual circle is parametrized by $\phi$. The T-dual geometry of the relevant coordinates is actually a trivial cartesian product $\IR^3\times \IS^1$, and the non-triviality of the Taub-NUT metric turns the presence of an NS5-brane located at a point\footnote{The position of the NS5-brane is controlled by the NSNS $B$-field along $\omega_2$ in the Taub-NUT picture. The physical gauge invariant quantity on the D7-brane worldvolume ${\cal {F}}_2=F_2-B_2$ is thus mapped to the distance between the NS5- and the D6-brane. For simplicity, we choose $B_2=0$, equivalently use the NS5-brane to define the origin in the $\IS^1$.} in the T-dual $\IS^1$. In the absence of the NS5-brane, $\phi$ would be a true Wilson line, with a continuous shift symmetry and unit period (in suitable units). The presence of the NS5-brane breaks the translational symmetry of $\IS^1$, and so also the shift symmetry of $\phi$. The non-trivial monodromy in $\phi$ arises because when the D6-brane moves around the $\IS^1$, it crosses the NS5-brane and there is a Hanany-Witten brane creation effect \cite{Hanany:1996ie} which creates a D4-brane stretched between the D6- and the NS5-branes, see Figure \ref{fig:hw}. As the axion continues completing turns around its period, more and more of these D4-branes are created. They contribute to the tension with a term which depends on the axion.

%%%%%%%%%%%
%%%%%%%%%%%

An interesting property manifest from the original and the T-dual pictures is that the energy increase due to the monodromy grows quadratically with the axion, {\em even for large fields}. This is due to the brane creation in each crossing in the T-dual picture, or to the appearance of induced D3-brane charge (which depends quadratically on $F_2$, even for large fields, as follows from the BPS bound) in the massive Wilson line picture. This scenario, if successfully embedded in a compact setup, may possibly lead to the construction of open axion monodromy models producing chaotic inflation, of the kind favored by recent observational data. We leave the issue of such embedding for future work.

\subsection{D-branes in type IIA and F-terms}

In this section we briefly discuss an example of a closed string axion model, whose F-term arises from D-brane couplings, rather than from fluxes. In a sense, it is an F-term version of the models in \cite{am1,am2}. It is based on type IIA orientifold compactifications with D6-branes, and with the axion $\phi$ arising from the NSNS B-field on a 2-cycle. The F-term is then generated from the coupling of $\phi$ to the D6-brane worldvolume. 

Explicitly, we expand B-field as $B = b\, \omega_b$,  with $\omega_b$ an harmonic integer orientifold even (1,1)-form (namely the imaginary part of a complexified K\"ahler modulus $T_b$). We also consider a D6-brane wrapping a 3-cycle $\Pi_3$, such that $\Pi_3$ has a non-trivial one-cycle. By Poincar\'e duality there is also a non-trivial two-cycle $\Sigma_2 \subset \Pi_3$, which may or may not be trivial in the compactification space $X_6$. If it is non-trivial then it may be that the pull-back of $\omega_b$ is also non-trivial in the homology of $\Pi_3$. Let us asume that this is the case and consider that 
\be
\int_{\Sigma_2} \omega_b = 1
\ee
We then have that a non-trivial superpotential is generated for the open string modulus $\Phi$ of the D6-brane
\be
W_{\rm D6} \sim \Phi \cdot T_b
\label{supob}
\ee
from which one could derive an effective F-term potential for the axion $b$. However, the potential derived from (\ref{supob}) is an approximation valid for small values of b. The complete expression, valid even for large field values is of the form
\be
V\, =\, \sqrt{({\rm Vol}_{\Pi_3})^2 + |T_b|^2} - {\rm Vol}_{\Pi_3}
\label{potF}
\ee
directly from  the DBI action, so for large values of $b$ we get a linear behaviour. 

One can show that for small values of $b$ one recovers (part of) the scalar potential related to (\ref{supob}). An example of this sort of computation can be seen in section 5.2 of \cite{fim06}, but with $\Phi T_b$ replaced by $\Phi(T_iT_j -n)$ in (\ref{supob}). Such kind of superpotentials are of the kind appearing for coisotropic D8-branes in type IIA compactifications, as well as D6-branes at angles T-dual to them. The latter also show that, by considering coistropic D8-branes, one may easily have potentials that behave like 
\be
V\, \sim\, b^2
\label{potFF}
\ee
for asymptotically large values of $b$. In both of these cases, D6-brane and coisotropic D8-branes, the presence of $b$ induces (non-conserved) D4-brane charge and tension, and so the reasoning of \cite{Conlon:2011qp} involving the backreaction of these objects would not apply. 

%These constructions can be used to construct generalizations of (\ref{potF}) for several axions. 

\section{Dimensional reduction and torsion}
\label{dimred}

\subsection{Torsional $p$-forms}
\label{dimtor}

Let us consider a manifold $\IX_d$ such that $ {\rm Tor} H^{p+1}(\IX_d,\IZ) \, =\, \IZ_k$, and a representative of the generator class of this torsion group. We have that
\beqa
d\eta_p=k\omega_{p+1}
\eeqa
for a globally well-defined $p$-form $\eta_p$. Let us further assume that $\omega_{p+1}$ is an eigenform of the Laplacian $\Delta = d d^* + d^*d$. Because $[\Delta, d] = 0$, $\eta_p$ is also an eigenform of the Laplacian. Namely we have that 
\be
\Delta \omega_{p+1} \, =\, - \mu^2 \, \omega_{p+1} \quad \quad \quad \Delta \eta_{p} \, =\, - \mu^2 \, \eta_{p}
\ee
We now reduce a potential $C_p$ along $\eta_p$ 
\be
C_p \, =\, \phi\, \eta_p \quad \quad \Raw \quad \quad F_{p+1}\, =\, \p_\mu \phi\, dx^\mu \wedge \eta_p + \phi \,k\,\omega_{p+1}
\ee
Then from a Lagrangian term containing $|F_{p+1}|^2$ we obtain
\be
\int_{\IX^{1,3}\times\IX_d}  F_{p+1} \wedge * F_{p+1}\, =\,  \int_{\IX^{1,3}} d\phi \wedge *_4 d\phi \, \int_{\IX_6} \eta_p \wedge * \eta_p +  \int_{\IX^{1,3}} *_4 (k \phi)^2 \, \int_{\IX_6} \omega_{p+1} \wedge * \omega_{p+1} 
\ee
Using the relations above we find 
\be
\int_{\IX_6} \omega_{p+1} \wedge * \omega_{p+1}\, =\, \frac{\mu^2}{{k^2}}  \int_{\IX_6}\eta_p \wedge * \eta_p
\ee
so we finally have
\be
\int_{\IX^{1,3}} \left(\p_\mu\phi\, \p^\mu \phi + {\mu^2} \phi^2 \right) d{\rm vol}_{X^{1,3}} \cdot  \int_{\IX_6} \eta_p \wedge * \eta_p
\ee
so if $\phi$ is canonically normalized $\mu$ is the mass of the axion. One can easily generalize this system for several massive axions whenever $ {\rm Tor} H^{p+1}(\IX_6,\IZ) \, =\, \IZ_{k_1} \oplus \IZ_{k_2} \oplus \dots \oplus \IZ_{k_n}$ along the lines of \cite{Camara:2011jg}.

Notice that for the case of massive Wilson line in $\tilde\IT^3$ discussed below we have
\be
\mu^2 \, =\, \frac{\int_{\tilde\T^3} d\eta^1 \wedge * d\eta^1}{\int_{\tilde\T^3} \eta^1 \wedge * \eta^1} \, =\, \left(\frac{kR_1}{R_5R_6}\right)^2
\ee
In general we expect $\mu \sim {\rm Vol} (\p \Sigma_p)/  {\rm Vol} (\Sigma_p)$ where $\Sigma_p$ is a $p$-chain related to the appropriate torsion class.

One can generalize the above to torsion and fluxes giving masses to RR axions by replacing $d \raw d_H = d - H \wedge$ everywhere. Now one has a linear combination of $p$-forms $\omega = \omega_{p+1} + \omega_{p+3} +\dots$ in $H^{p+1}(\IX_6,\IZ) \oplus H^{p+3}(\IX_6,\IZ) \oplus \dots$ such that
\beqa
d_H\eta=k\omega
\eeqa
such that $\eta$ is another polyform.\footnote{Exact polyforms $k' \eta = d_H \tilde{\omega}$ are eaten by some $U(1)$ and should be removed from the spectrum before starting this discussion.}
We also require that $\omega$ is an eigenvalue of the Laplacian $\Delta_H$ constructed from $H$, and then we have the same effective action with
\be
\mu^2 \, =\, \frac{\int_{X_6} d_H\eta \wedge * d_H\eta}{\int_{X_6} \eta \wedge * \eta} 
\ee

\subsection{Massive Wilson lines}
\label{dimDBI}

Let us consider a twisted torus geometry ${\bf \tilde{T}^6}$ of the form
\be
ds^2 =  R_1^2(dx^1 -k\, x^6 dx^5)^2 + R_2^2(dx^2 + k\, x^4 dx^6)^2 +  \sum_{i=3}^6 (R_i^2dx^i)^2 
\label{metricIIAap}
\ee
with a fundamental form given by
\be
J\, =\, R_1R_4\, \eta^1 \wedge \eta^4 + R_2R_5\, \eta^2 \wedge \eta^5 + R_3R_6\, \eta^3 \wedge \eta^6
\ee
where the 1-forms $\eta^i$ are globally well-defined and eigenforms of the laplacian $\Delta = dd^* + d^*d$, namely
\be
\begin{array}{ccc}
\eta^1 = dx^1 - k\, x^6 dx^5 & \quad & \eta^4 = dx^4 \\
\eta^2 = dx^2 + k\, x^4 dx^6 & \quad & \eta^5 = dx^5 \\
\eta^3 = dx^3 & \quad & \eta^6 = dx^6 
\end{array}
\label{etas}
\ee
We now consider a D6-brane wrapping the submanifold $\pi_3 = \{x^2=x^3=x^4=0\}$, so that its volume form is given by $d{\rm vol}_{\pi_3} = R_1R_5R_6 \eta^1 \wedge \eta^5 \wedge \eta^6$. Since $J|_{\pi_3} = 0$ this is indeed a Lagrangian submanifold of ${\bf \tilde{T}^6}$. We now turn on a Wilson line along the direction $\eta^1$
\be
A\, =\, \left( \langle w\rangle + w \right)\, \eta^1
\ee
because $d\eta^1 = kdx^5 \wedge dx^6$ a non-vanishing vev for $w$ will induce a flux $F=dA$ on the D6-brane worldvolume, so this will be a massive Wilson line. 

In order to see how the energy of the system changes, let us consider the DBI action of the D6-brane on $\pi_3$. The worldvolume metric and field strength are given by
\bea
ds^2 & = & ds^2_{X^{1,3}} + R_1^2(dx^1 -k\, x^6 dx^5)^2 + (R_5 dx^5)^2  + (R_6 dx^6)^2 \\
F & = & \p_\mu w\, dx^\mu\wedge \eta^1 + {k} \left( \langle w\rangle + w \right) dx^5 \wedge dx^6
\eea
we then have that the DBI action reads
\be
S_{DBI}\, =\, -\mu_6  \int d^7\xi e^{-\phi} \sqrt{ {\rm det\, } (G + 2\pi \a' F)}
\ee
and it is easy to see that the square root reads
\be
\sqrt{ {\rm det\, } (G + 2\pi \a' F)}\,=\, d{\rm vol}_{X^{1,3}}\, \sqrt{\left(R_1^2+ \p_\mu w\p^\mu w\right) \left(R_5^2R_6^2 + (2\pi\a')^2 k^2\langle w\rangle^2 \right)} 
\ee
and so the kinetic terms and D-brane energy read
\be
S_{DBI} \, = \, - \mu_6g_s^{-1}   \int_{X^{1,3}} \left(*_41\cdot V_6 + \frac{1}{2}\frac{\tilde{V}_6}{R_1^2} dw \wedge *_4 d{w} \right)
\ee
where
\be
\tilde{V}_6(\langle w \rangle)\, =\, \int_{\pi_3} \sqrt{1+ \frac{(2\pi \a')^2}{R_5^2R_6^2} k^2 \langle w \rangle^2 }  d{\rm vol}_{\pi_3}\, =\, R_1 \sqrt{R_5^2R_6^2 + (2\pi\a')^2 k^2\langle w\rangle^2} 
\ee
and
\be
{V}_6(\langle w \rangle)\, =\, \tilde{V}_6(\langle w \rangle) - R_1R_5R_6
\ee
of course one should express this action in the Einstein frame, which basically amounts to divide by ${\rm Vol}_{\bf \tilde{T}_6}$. Notice that in the limit
\be
\langle w \rangle \ll \frac{R_5R_6}{2\pi k \a'}
\ee
we can neglect the Wilson line dependence in $V_6$ and we have that the kinetic term is proportional to 
\be
\frac{\tilde{V}_6}{R_1^2} \sim \frac{R_5R_6}{R_1} = \int_{\pi_3} \eta^1 \wedge *_3 \eta^1 \equiv {\cal H}
\ee
as expected from the case of massless Wilson lines (\ref{H}). For large values of $\langle w \rangle$ this dependence cannot be neglected.

\end{document}